\documentclass[english,pra, twocolumn]{revtex4-1}
\usepackage[T1]{fontenc}
\usepackage[latin9]{inputenc}
\usepackage{color}
\definecolor{note_fontcolor}{rgb}{0.80078125, 0.80078125, 0.80078125}
\usepackage{babel}
\usepackage{amsmath}
\usepackage{amssymb}
\usepackage{graphicx}
\usepackage{esint}
\usepackage[unicode=true,pdfusetitle,
 bookmarks=true,bookmarksnumbered=false,bookmarksopen=false,
 breaklinks=false,pdfborder={0 0 1},backref=false,colorlinks=false]
 {hyperref}

\makeatletter

\newenvironment{lyxgreyedout}
  {\textcolor{note_fontcolor}\bgroup\ignorespaces}
  {\ignorespacesafterend\egroup}

 \@ifundefined{textcolor}{}
 {%
   \definecolor{BLACK}{gray}{0}
   \definecolor{WHITE}{gray}{1}
   \definecolor{RED}{rgb}{1,0,0}
   \definecolor{GREEN}{rgb}{0,1,0}
   \definecolor{BLUE}{rgb}{0,0,1}
   \definecolor{CYAN}{cmyk}{1,0,0,0}
   \definecolor{MAGENTA}{cmyk}{0,1,0,0}
   \definecolor{YELLOW}{cmyk}{0,0,1,0}
 }

\makeatother

\begin{document}

\title{Three-Dimensional Spin-Orbit Coupling in a Trap}

\author{Brandon M. Anderson and Charles W. Clark}

\affiliation{Joint Quantum Institute, National Institute of Standards and Technology
and the University of Maryland, Gaithersburg, Maryland, 20899-8410,
USA}
\begin{abstract}
We investigate the properties of an atom under the influence of a
synthetic three-dimensional spin-orbit coupling (Weyl coupling) in
the presence of a harmonic trap. The conservation of total angular
momentum provides a numerically efficient scheme for finding the spectrum
and eigenfunctions of the system. We show that at large spin-orbit
coupling the system undergoes dimensional reduction from three to
one dimension at low energies, and the spectrum is approximately Landau
level-like. At high energies, the spectrum is approximately given
by the three-dimensional isotropic harmonic oscillator. We explore
the properties of the ground state in both position and momentum space.
We find the ground state has spin textures with oscillations set by
the spin-orbit length scale.
\end{abstract}
\maketitle
\begin{lyxgreyedout}
\global\long\def\ket#1{\left|#1\right\rangle }

\global\long\def\bra#1{\left\langle #1\right|}

\global\long\def\onehalf{\frac{1}{2}}

\global\long\def\atwo{\frac{\lambda}{2}}

\global\long\def\Op{\mathcal{A}^{+}}

\global\long\def\Om{\mathcal{A}^{-}}
\end{lyxgreyedout}

\section{Introduction}

The recent experimental success in simulating spin-orbit coupling\cite{SpielmanMag,SpielmanPRL,SpielmanSO,SpielmanE,BlochMag}
in a cold atom context has produced great interest in the field. Although
the experimental setup produced only an Abelian spin-orbit coupling,
it is possible to produce more complicated non-Abelian synthetic fields,
such as any combination of Rashba and linear Dresselhaus couplings.\cite{Nlevel,NonAbelian,Npod,GaugeReview,SOBEC,SpinOrbit}
The isotropic limit, such as Rashba or linear Dresselhaus coupling,
is particularly interesting. In this limit, and the absence of an
external potential, the ground state of the spin-orbit coupling has
an infinitely degenerate ground state manifold, which is given by
a ring in momentum space. The ring of minima gives an energetically
free direction for low-energy quantum fluctuations. This leads to
an effective dimensional reduction for low energy properties\cite{TrappedRashbaHui,TrappedRashbaHuiLong,TrappedRashbaSantos},
such as a Landau-like spectrum \cite{ShenoyTrappedSO,3DTI.Wu,TrappedRashbaSantos,TrappedRashbaHui},
enhanced binding energy \cite{BS1,BS2,Rashbon,Scooped,SoBoundStates,BoundStates1,BoundStates2,BoundStates3,RashbaBECBCS.Shenoy},
or spontaneous symmetry breaking\cite{RashbaOrderByDisorder}. In
the presence of a trap or interactions this degeneracy will be broken,
up to the two-fold degeneracy guaranteed by time-reversal symmetry.
However, as we show here, effects of the infinite manifold of states
will survive even to the trapped regime, and the dimensional reduction
of low energy states will still be visible.

A recent proposal \cite{3DSOC,Goldman,3DTI.Wu} allows for the study
of spin-orbit coupling that has no solid state counterpart: three-dimensional
spin-orbit coupling, or Weyl coupling\cite{WeylFermions}, of the
form: $v\mathbf{p}\cdot\boldsymbol{\sigma}$, where $\mathbf{p}$
is momentum, and $\boldsymbol{\sigma}$ is the spin operator. The
conceptual picture in this system is similar to that of the Rashba
case. However, instead of a ring of ground states we now have a spherical
manifold of ground states. This implies there are two energetically
free directions for low energy quantum fluctuations, so the dimensional
reduction is now from $D=3$ to $D=1$. 

In this paper we analyze the motion of a particle moving in an isotropic
harmonic trap, subject to Weyl coupling. We show that while Weyl coupling
is the natural extension of Rashba coupling to incorporate the third
spatial dimension, this extension actually increases the overall symmetry
of the system and simplifies its treatment. We describe a numerical
scheme for finding the eigenvalues exactly, and we analyze the eigenfunctions.
Even the ground state of this system exhibits significant spin and
orbital textures. In the limit of large coupling we find Landau like
behavior of the spectrum, in that the low-energy spectrum tends to
that of a one-dimensional harmonic oscillator with a large number
of nearly degenerate levels. However, at sufficiently large energies,
the spin-orbit coupling becomes a perturbation, and the high energy
spectrum is well approximated by the 3D isotropic harmonic oscillator.

The paper is organized as follows. We first review the model Weyl
coupling Hamiltonian. We perform a partial wave expansion of the system
in the presence of a trap, and find a numerically efficient method
for diagonalizing the system. We calculate the spectrum as a function
of the spin-orbit coupling strength, and find the low energy spectrum
becomes Landau-like at large coupling. At sufficiently high energies,
the spectrum crosses over to that of the three-dimensional isotropic
harmonic oscillator. We derive a system of coupled differential equations
in the radial coordinate only, and use it to explain the crossover
between the Landau spectrum and isotropic oscillator spectrum. Finally,
we show the ground state is characterized by persistent orbital and
spin currents.

\section{Model}

Our Hamiltonian is given by 
\begin{equation}
H=\frac{{\bf p}^{2}}{2}+v{\bf p}\cdot\boldsymbol{\sigma}+\frac{{\bf r}^{2}}{2},
\end{equation}
where $\mathbf{r}$ and $\mathbf{p}$ are respectively the position
and momentum operators and $\boldsymbol{\sigma}=\left(\sigma_{1},\sigma_{2},\sigma_{3}\right)$
is the vector of conventional Pauli matrices. This Hamiltonian can
be implemented in cold atomic systems using two-photon transitions
as described in \cite{3DSOC}: here $v$ is a coupling constant subject
to external control. The two dimensions of the spin operator $\boldsymbol{\sigma}$
correspond to two internal states of the atom, which would ordinarily
be two different hyperfine states. We have chosen a system of units
in which the reduced Planck constant $\hbar$, the mass $M$ of the
particle, and the harmonic oscillator frequency $\omega$ are all
equal to 1. 

Our Hamiltonian commutes with an angular momentum $\mathbf{J}=\mathbf{L}+\frac{\boldsymbol{\sigma}}{2}$,
where \textbf{$\mathbf{L}=\mathbf{r}\times\mathbf{p}$} is the orbital
angular momentum of the atomic center of mass. Note that this $\mathbf{J}$
need not be identical to the usual angular momentum that is constructed
from the sum of the orbital angular momentum of the atom\textquoteright{}s
center of mass and the angular momenta of its electrons and nuclei.
However, it is a conserved quantity whose three spatial components
satisfy the usual commutation relations of an angular momentum operator,
and in this sense we can treat $\mathbf{J}$ as the sum of an integer-valued
orbital angular momentum\textbf{ $\mathbf{L}$} and a spin $s=\frac{1}{2}$.
The $\boldsymbol{\sigma}\cdot\mathbf{p}$ term is a scalar under the
rotations induced by $\mathbf{J}$, as is the kinetic and trapping
term, so our Hamiltonian commutes with $\mathbf{J}$. We note in passing
that Rashba coupling is obtained from removing the $\sigma_{3}p_{3}$
term from Weyl coupling. Thus Rashba coupling is a linear combination
of scalar and rank-2 tensor operators under rotations, whereas Weyl
coupling has the simpler scalar form and is much simpler to solve
with a standard partial-wave decomposition. 

In the absence of a trap, the spectrum of the Weyl coupling can be
found exactly. There are two bands corresponding to spin aligned,
and anti-aligned with momentum. The spectrum is given by 
\begin{equation}
E(\mathbf{p})=\frac{\mathbf{p}^{2}}{2}\pm v|\mathbf{p}|.
\end{equation}
The low energy band has a minimum defined on the sphere $|\mathbf{p}|=v$.
Near the minimum of this sphere, the dispersion is parabolic only
along the radial direction, and is constant along the polar and azimuthal
axes. This suggests that for low energy properties of the system,
quantum fluctuations will be energetic only along the radial direction.
We expect that in the presence of a spherically symmetric trap, the
low energy spectrum of the system will undergo a dimensional reduction
from $D=3$ to $D=1$ for a sufficiently large spin-orbit coupling. 

 The conserved total angular momentum $\mathbf{J}$ allows us to use
a basis of well defined angular momentum $|j,m,\lambda\rangle$, where
\begin{equation}
\ket{j,m,\lambda}=e^{i\phi_{\lambda}}\begin{pmatrix}\lambda\sqrt{\frac{j-\lambda m+x_{\lambda}}{2(j+x_{\lambda})}}\ket{j+\atwo,m-\onehalf}\\
-\sqrt{\frac{j+\lambda m+x_{\lambda}}{2(j+x_{\lambda})}}\ket{j+\atwo,m+\onehalf}
\end{pmatrix}\label{eq:jmlambda}
\end{equation}
and $\lambda=\pm1$ corresponds to ${\bf J}$ and $\boldsymbol{\sigma}$
aligned $(j=l+s)$ or anti-aligned $(j=l-s)$. These states have eigenvalues
${\bf J}^{2}|j,m,\lambda\rangle=j(j+1)|j,m,\lambda\rangle$ and $J_{z}|j,m,\lambda\rangle=m|j,m,\lambda\rangle$.
They are complete, which allows us to project our Hamiltonian into
subspaces of fixed $j,m$.

\subsection{Number Basis}

The Weyl coupling Hamiltonian can be numerically diagonalized by performing
a partial wave decomposition into states of the three-dimensional
isotropic harmonic oscillator, with angular momentum state $\ket{j,m,\lambda}$.
We define the basis $\ket{n,j,m,\lambda}=\ket n\ket{j,m,\lambda}$
in Appendix \ref{sec:Appendix-A} as a state with $n$ radial quantum
nodes, and an angular momentum eigenstate given by (\ref{eq:jmlambda}).
When $v=0$ these states have energy $E=2n+l+\frac{3+\lambda}{2}$.
\cite{3DSHO_Basis} It is the convenient to express the spin-orbit
coupling in terms of creation(annihilation) operators $\mathbf{a^{\dagger}}(\mathbf{a})$
as 
\begin{eqnarray}
{\bf p}\cdot\boldsymbol{\sigma} & = & \frac{-i}{\sqrt{2}}\left(\boldsymbol{\sigma}\cdot{\bf a}^{\dagger}-\boldsymbol{\sigma}\cdot{\bf a}\right)\\
 & = & \frac{-i}{\sqrt{2}}\left(\Op-\Om\right)
\end{eqnarray}
where $\Op=\boldsymbol{\sigma}\cdot{\bf a}^{\dagger}$ and $\Om=\boldsymbol{\sigma}\cdot{\bf a}$
are two rank-0 tensors with respect to rotations generated by $\mathbf{J}$.
These operators satisfy the commutation relation $\frac{1}{2}\left\{ \Op,\Om\right\} =\mathbf{a}^{\dagger}\cdot\mathbf{a}+\frac{3}{2}=E/\hbar\omega$,
which allows us to express our spin-orbit coupled Hamiltonian as
\begin{equation}
H=\frac{1}{2}\left\{ \Op,\Om\right\} +v\frac{i}{\sqrt{2}}\left(\Op-\Om\right).
\end{equation}
In Appendix \ref{sec:Appendix-A}, we show that the matrix elements
of the operators $\Op$ and $\Om$ are given by 
\begin{eqnarray}
\Om|n,j,m,-\rangle & = & \sqrt{2(n+j+1)}|n,j,m,+\rangle\label{eq:Aopm}\\
\Om|n,j,m,+\rangle & = & \sqrt{2n}|n-1,j,m,-\rangle\nonumber 
\end{eqnarray}
and
\begin{eqnarray}
\Op|n,j,m,+\rangle & = & \sqrt{2(n+j+1)}|n,j,m,-\rangle\label{eq:Aopp}\\
\Op|n,j,m,-\rangle & = & \sqrt{2(n+1)}|n+1,j,m,+\rangle.\nonumber 
\end{eqnarray}

Since the operator $\Op$ contains a combination of creation operators,
it is clear that it will raise the energy of a state by one unit.
In the radial basis there are two ways to raise the energy by one
unit. The angular quantum number can be increased by one with the
radial quantum number held constant: $\Delta l=+1$, $\Delta n=0$,
or the radial number can be increases by one and the angular quantum
number lowered by one. Repeated applications of the raising operator
alternate between state with $l=j+s$ and $l=j-s$, while increasing
the energy by one unit each time.

\subsection{Numerical Diagonalization\label{sub:Numerical-Diagonalization}}

\begin{figure}
\includegraphics[width=0.9\columnwidth]{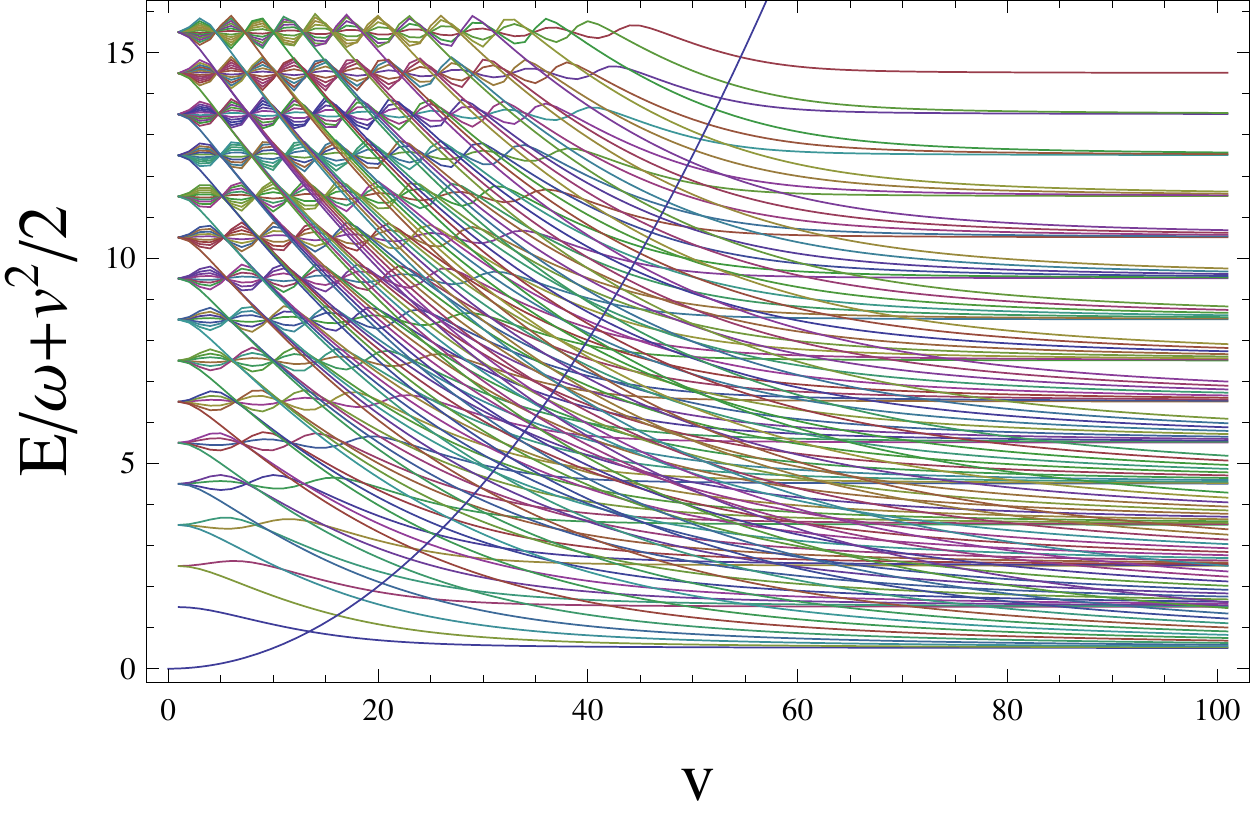}

\caption{Numerical calculation of the spectrum as a function of $v$. At $v=0$,
all energy levels are plotted have energy $E\leq15+\frac{3}{2}$.
As the spin-orbit strength is increased, the levels split off into
groups with increasing radial quantum number $n$. Each landau level
has energy of approximately $E_{njm}=\left(n+\frac{1}{2}\right)+\frac{\left(j+\frac{1}{2}\right)^{2}}{2v^{2}}-\frac{v^{2}}{2}$.
As a visual guide, each level is shifted by $v^{2}/2$. Two regimes
are clearly identifiable, corresponding to a $3D$ harmonic oscillator
with slight level mixing, and the $1D$ Landau-level problem. As discussed
in the text, the crossover between these two regimes is given by $E\sim v^{2}/2$.
This crossover is shown by the black line.}
\end{figure}

The spin-orbit coupled Hamiltonian can therefore be numerically diagonalized
efficiently by first projecting the Hamiltonian into sectors of good
$|j,m\rangle$
\begin{equation}
H=\sum_{j}\sum_{m=-j}^{j}H_{jm}|jm\rangle\langle jm|
\end{equation}
where the Hamiltonian $H_{jm}$ is defined by the matrix elements
\begin{eqnarray}
\langle n^{\prime}\lambda^{\prime}|H_{jm}|n\lambda\rangle & = & (2n+j+1)\delta_{n,n^{\prime}}\delta_{\lambda,\lambda^{\prime}}\\
 & + & iv\left(\sqrt{n+1}\delta_{n+1,n^{\prime}}\delta_{\lambda^{\prime}+}\delta_{\lambda-}\right.\nonumber \\
 &  & \quad\left.-\sqrt{n}\delta_{n-1,n^{\prime}}\delta_{\lambda+}\delta_{\lambda^{\prime}-}\right)\nonumber \\
 & + & iv\sqrt{n+j+1}\delta_{n,n^{\prime}}\left(\delta_{\lambda+}\delta_{\lambda^{\prime}-}-\delta_{\lambda^{\prime}+}\delta_{\lambda-}\right).\nonumber 
\end{eqnarray}
This matrix is tridiagonal, and can be efficiently diagonalized through
$\mathcal{O}(n)$ operations. Figure 1 shows the spectrum as a function
of the spin-orbit parameter $v$. At $v=0$, we have included all
states that have $N\leq10$, where $N=2n+l$ is the total quanta of
the 3D isotropic oscillator. Two regions in the spectrum are identifiable.
For $N\gg\frac{v^{2}}{2}$ the spectrum is approximately that of the
three-dimensional harmonic oscillator with $E\approx2n+l+\frac{3}{2}-\frac{v^{2}}{2}$.
For $N\ll\frac{v^{2}}{2}$ the spectrum is given by 
\begin{eqnarray}
E & \approx & \left(n+\frac{1}{2}\right)+\frac{\left(j+\frac{1}{2}\right)^{2}}{2v^{2}}-\frac{v^{2}}{2}\label{eq:Spectrum1}
\end{eqnarray}
as will be shown in Sec.\ref{sub:Landau-Levels}. To lowest order
in $1/v$, this has the form of a one dimensional harmonic oscillator
in the radial mode. We will later see that these states are well localized
in the momentum space potential near $|\mathbf{p}|=v$. This suggests
that the energetically free excitations along the polar and azimuthal
directions are not important for low energy states. The system therefore
undergoes a dimensional reduction from $D=3$ to $D=1$. The one-dimensional
structure is reminiscent of the Landau levels. Since the problem is
spherically symmetric, mixing of the Landau levels appears only at
higher order in the inverse spin-orbit coupling parameter. Note that
while these levels have been seen in previous work, \cite{ShenoyTrappedSO,3DTI.Wu,TrappedRashbaSantos,TrappedRashbaHui},
the crossover to the three-dimensional spectrum was missed.

\subsection{The Schr\"{o}dinger Equation as a system of coupled differential
equations}

\begin{figure}
\includegraphics[width=0.9\columnwidth]{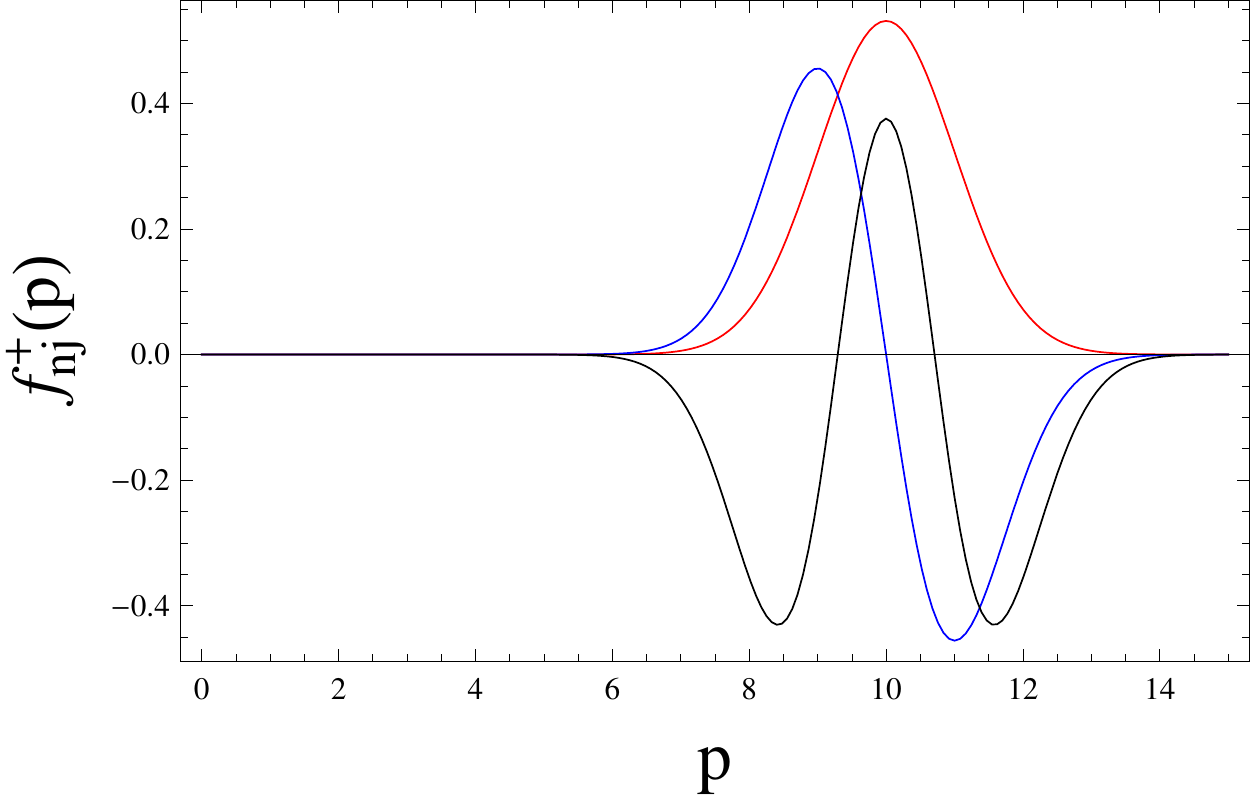}\caption{Momentum space eigenfunctions $f_{nj}^{+}(p)$ for low energy states
at large spin-orbit coupling. The red, blue and black curves correspond
to states with $n=1,2,3$ respectively. The eigenfunctions are well
approximated by the one dimensional harmonic oscillator wavefunction
centered around $p=v$. The number of nodes corresponds to the radial
quantum number. States with higher $j$ have the same form, up to
a possible sign, provided the number of radial quanta are smaller
than $\sim v^{2}/2$. The eigenfunctions $f_{nj}^{-}(p)$ have opposite
sign.}
\label{fig:MomEfuncts}
\end{figure}

\subsubsection{Momentum Space}

The eigenstates of the Hamiltonians $H_{jm}$ are states of good total
angular momentum. In general they can be expressed as 
\begin{equation}
\ket{n_{r},j,m}=\ket{\psi_{n_{r}}^{+}}\ket{j,m,+}+\ket{\psi_{n_{r}}^{-}}\ket{j,m,-},
\end{equation}
where $\ket{\psi_{n_{r}}^{\pm}}$ is an eigenstate of $H_{jm}$ with
$n_{r}$ radial modes. This form suggests that each Hamiltonian $H_{jm}$
has a corresponding set differential equation in only the radial degrees
of freedom. It is more natural to work in momentum space, where $\mathbf{p}$
is a dynamical variable, instead of an operator. The presence of the
harmonic trap allows us to treat the position operator as a derivative,
$\mathbf{r}=i\nabla_{\mathbf{p}}$. The corresponding ``dual'' Schr\"{o}dinger
equation is
\begin{equation}
\left[-\frac{\nabla_{\mathbf{p}}^{2}}{2}+\frac{\mathbf{p}^{2}}{2}+v\boldsymbol{\sigma}\cdot\mathbf{p}\right]\psi(\mathbf{p})=E\psi(\mathbf{p}).
\end{equation}
The radial eigenfunctions in momentum space have the corresponding
form
\begin{equation}
\psi_{njm}(\mathbf{p})=f_{nj}^{-}(p)\chi_{jm}^{-}(\hat{\mathbf{p}})+f_{nj}^{+}(p)\chi_{jm}^{+}(\hat{\mathbf{p}}),\label{eq:RadialAnsatz}
\end{equation}
where $p=|\mathbf{p}|$, $\hat{\mathbf{p}}=\mathbf{p}/p$, $f_{nj}^{\pm}(p)=\langle p|\psi_{n_{r}}^{\pm}\rangle$,
and the spinors $\chi_{jm}^{\pm}=\langle\hat{\mathbf{p}}|jm\pm\rangle$
have total momentum $j$, with a $J_{3}$ projection $m$. As shown
in the appendix, the action of the operator $\boldsymbol{\sigma}\cdot\mathbf{p}$
operating on the spinors $\chi_{jm}^{\pm}(\hat{\mathbf{p}})$ is to
interchange the spinors and multiply the result by $p$, i.e., $\boldsymbol{\sigma}\cdot\mathbf{p}\chi_{jm}^{\pm}(\hat{\mathbf{p}})=p\chi_{jm}^{\mp}(\hat{\mathbf{p}})$.
The form (\ref{eq:RadialAnsatz}) gives a consistent set of two coupled
differential equations in the independent variable $p$. These coupled
differential equations takes the form\begin{widetext}
\begin{eqnarray}
\left[-\frac{1}{2}\frac{\partial^{2}}{\partial p^{2}}+\frac{1}{2}\frac{\left(j+\frac{1}{2}\right)\left(j+\frac{3}{2}\right)}{p^{2}}+\frac{p^{2}}{2}\right]u_{nj}^{+}(p)+v\, pu_{nj}^{-}(p) & = & E\, u_{nj}^{+}(p)\label{eq:RadMomDEQ1}\\
\left[-\frac{1}{2}\frac{\partial^{2}}{\partial p^{2}}+\frac{1}{2}\frac{\left(j-\frac{1}{2}\right)\left(j+\frac{1}{2}\right)}{p^{2}}+\frac{p^{2}}{2}\right]u_{nj}^{-}(p)+v\, pu_{nj}^{+}(p) & = & E\, u_{nj}^{-}(p)\label{eq:RadMomDEQ2}
\end{eqnarray}
\end{widetext}where $u_{nj}^{\pm}(p)=p\, f_{nj}^{\pm}(p)$. 

The eigenfunctions $f_{nj}^{\pm}(p)$ can be found by solving these
coupled differential equations. Alternatively, they can be constructed
using the eigenvectors found from diagonalizing the projected Hamiltonians
$H_{jm}$. Some examples of these functions are shown in Fig.\ref{fig:MomEfuncts}
for a large $v$. The functions are well approximated by the one-dimensional
harmonic oscillator wavefunctions centered around $p=v$.

\subsubsection{Position space}

\begin{figure}
(a) Ground State

\includegraphics[width=0.9\columnwidth]{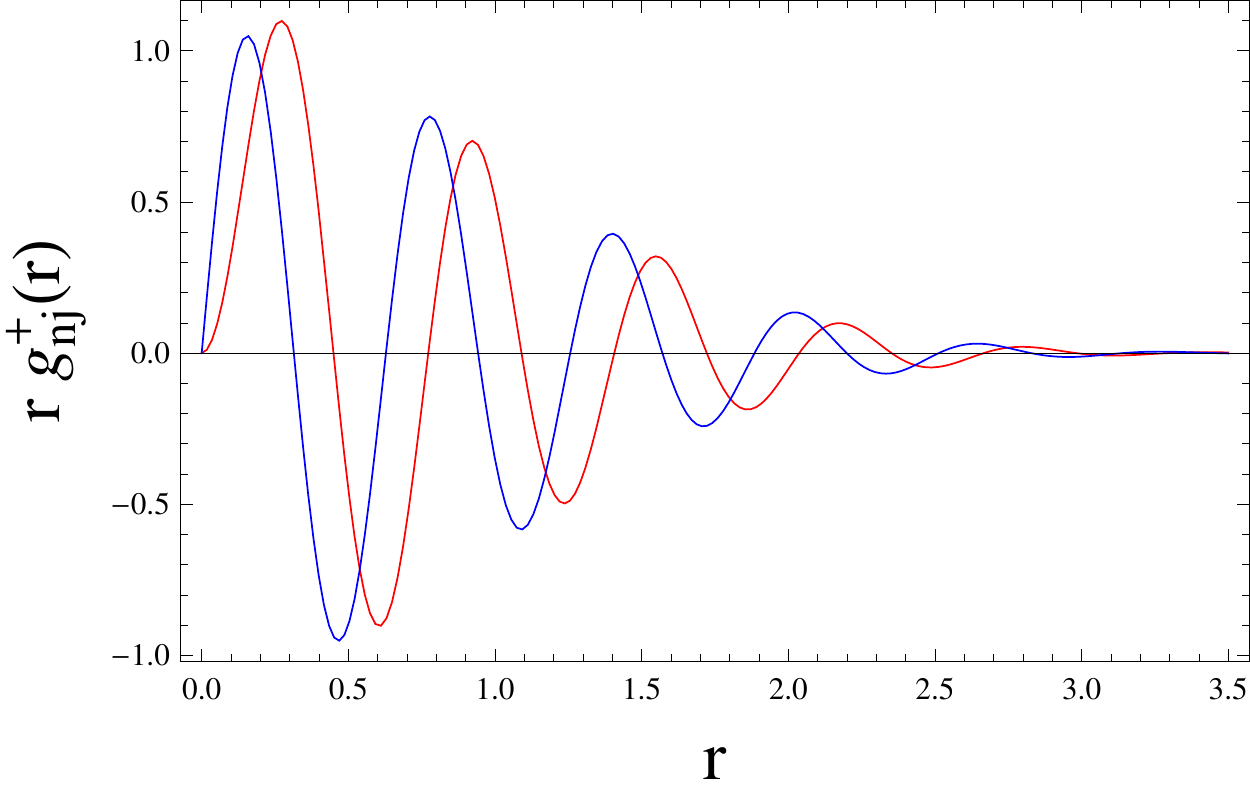}

(b) Lowest radial numbers

\includegraphics[width=0.9\columnwidth]{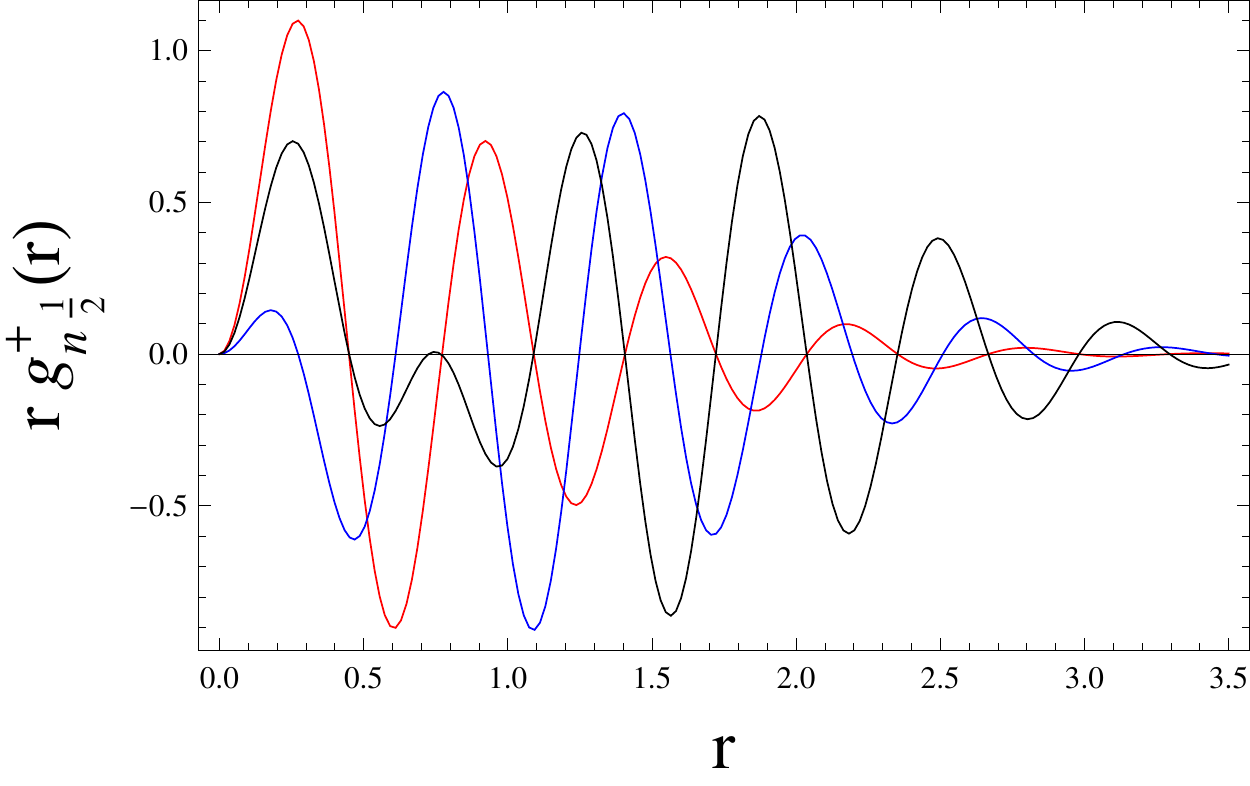}

(c) Lowest $j$ values

\includegraphics[width=0.9\columnwidth]{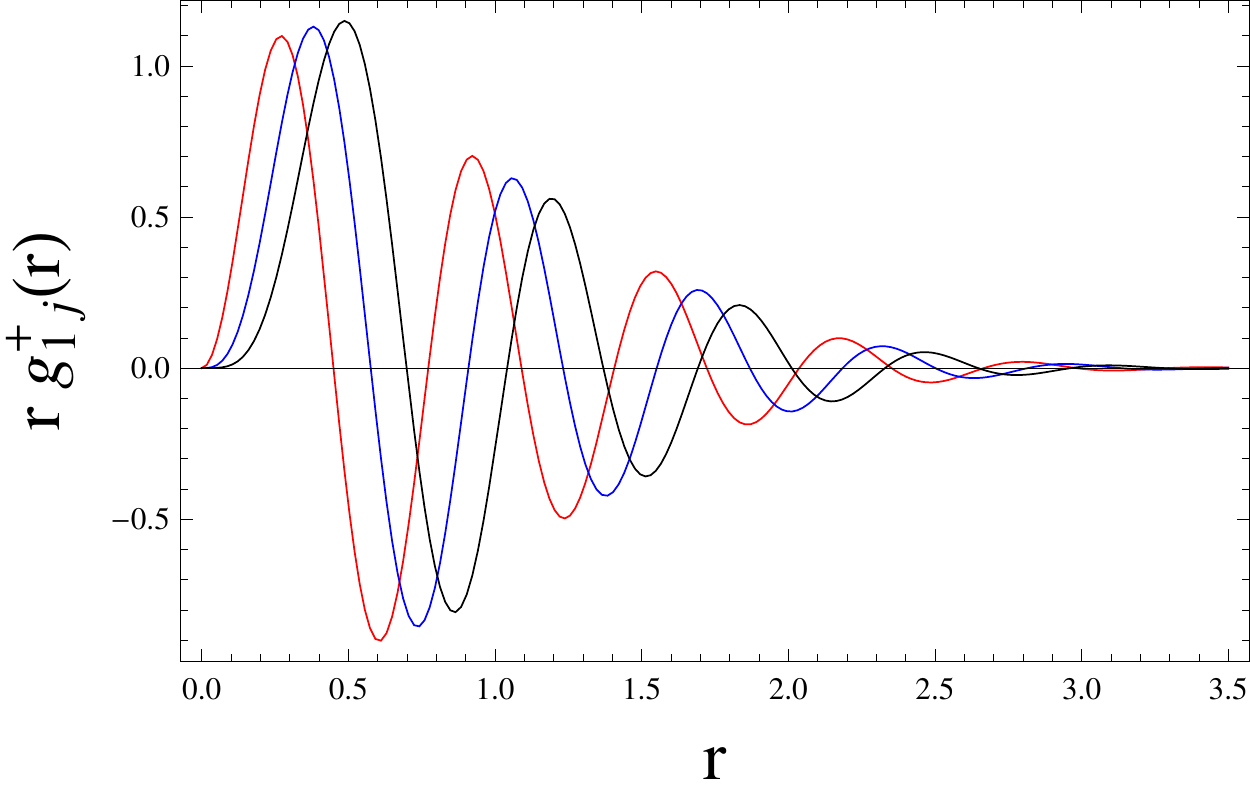}

\caption{(a) Position space eigenfunctions for low ground states at large spin-orbit
coupling. The red and blue curves correspond to the functions $rg_{nj}^{+}(r)$
and $rg_{nj}^{-}(r)$ respectively. (b) The first three position space
eigenfunctions $rg_{nj}^{+}(r)$ for $j=1/2$ and $n=1,2,3$$ $ (red,
blue and black respectively.) (c) The position space eigenfunctions
$rg_{nj}^{+}(r)$ in the lowest radial mode for three lowest values
of $j=\frac{1}{2},\frac{3}{2},\frac{5}{2}$, correspond to the red,
blue and black curves respectively. }
\label{fig:PosEfuncts}
\end{figure}
A radial differential equation can be found in position space in a
manner analogous to the momentum space differential equations. The
position space analogue of (\ref{eq:RadialAnsatz}) is given by 
\begin{equation}
\psi_{njm}(\mathbf{r})=g_{nj}^{-}(r)\chi_{jm}^{-}(\hat{\mathbf{r}})+g_{nj}^{+}(r)\chi_{jm}^{+}(\hat{\mathbf{r}}),\label{eq:RadialAnsatzPosition}
\end{equation}
where $r=|\mathbf{r}|$, $\hat{\mathbf{r}}=\mathbf{r}/r$, $g_{nj}^{\pm}(r)=\langle r|\psi_{nj}^{\pm}\rangle$
and $\chi_{jm}^{\pm}(\mathbf{\hat{r}})=\langle\hat{\mathbf{r}}|jm\lambda\rangle$.
In the appendix we show that the functions $g_{nj}^{\pm}(r)$ satisfy
the radial differential equation\begin{widetext}
\begin{eqnarray}
\frac{1}{2}\left[-\frac{1}{r}\left(\frac{d^{2}}{dr^{2}}r\right)+\frac{(j+\frac{1}{2})(j+\frac{3}{2})}{r^{2}}+r^{2}\right]g_{nj}^{+}(r)+v\left(-\frac{d}{dr}+\frac{j-\frac{1}{2}}{r}\right)g_{nj}^{-}(r) & = & E_{nj}g_{nj}^{+}(r)\\
\frac{1}{2}\left[-\frac{1}{r}\left(\frac{d^{2}}{dr^{2}}r\right)+\frac{(j+\frac{1}{2})(j-\frac{1}{2})}{r^{2}}+r^{2}\right]g_{nj}^{-}(r)+v\left(\frac{d}{dr}+\frac{j+\frac{1}{2}}{r}\right)g_{nj}^{+}(r) & = & E_{nj}g_{nj}^{-}(r).
\end{eqnarray}
\end{widetext}These differential equations can be solved to find
the eigenfunctions in position space. Alternatively, we can the radial
eigenfunction from the partial wave expansion found in Sec.\ref{sub:Numerical-Diagonalization}.

\subsubsection{Landau Levels\label{sub:Landau-Levels}}

The Landau levels we found numerically, which are described by (\ref{eq:Spectrum1}),
can be understood by considering the asymptotic form of the radial
differential equations, (\ref{eq:RadMomDEQ1}) and (\ref{eq:RadMomDEQ2}),
in momentum space. At large momentum, $p\gtrsim v$, the $1/p^{2}$
term becomes negligible, and the differential equations can be decoupled
by taking an even/odd superposition of (\ref{eq:RadMomDEQ1}) and
(\ref{eq:RadMomDEQ2}). The two decoupled differential equations are
\begin{eqnarray}
\left[-\frac{1}{2}\frac{\partial^{2}}{\partial p^{2}}+\frac{p^{2}}{2}\pm v\, p\right]\tilde{u}_{nj}^{\pm}(p) & = & E\,\tilde{u}_{nj}^{\pm}(p),\label{eq:RadMomDEQ1-1}
\end{eqnarray}
where $\tilde{u}^{\pm}=\frac{1}{2}\left(u^{+}\pm u^{-}\right)$. These
differential equations can be mapped to the one-dimensional harmonic
oscillator by performing a change of variables $p\rightarrow p\mp v$.
The solutions are given by $\tilde{u}_{nj}^{\pm}(p)=c_{n}H_{n}(p\pm v)e^{-\frac{(p\pm v)^{2}}{2}}$,
where $H_{n}(p)$ is the $n$-th Hermite polynomial, \cite{DLMF_Hermite}
and $c_{n}$ is a normalization constant. However, the equation for
$\tilde{u}^{+}(p)$ is localized around $p=-v$, outside of the range
of definition of the radial coordinate. We therefore assume these
solutions are $\tilde{u}_{nj}^{+}(p)=0$. Transforming, we find that
$u^{+}(p)=-u^{-}(p)=\sqrt{\frac{1}{2^{n}n!}}\frac{1}{\pi^{1/4}}H_{n}\left(p-v\right)e^{-\left(p-v\right)^{2}/2}$
in the asymptotic limit. 

The form of the harmonic oscillator Hamiltonian suggests the spectrum
is 
\begin{equation}
E_{n}=n+\frac{1}{2}-\frac{v^{2}}{2}.
\end{equation}
To lowest order in $v$, this result is consistent with the large
Landau-level like degeneracy found earlier. This degeneracy is lifted
by the centrifugal barrier, which mixes the Landau levels near $p=0$.
To lowest order in perturbation theory, the energy shift in the state
$\psi_{njm}$ is given by
\begin{eqnarray}
\delta E_{njm} & = & \langle\psi_{njm}|\frac{1}{2}\frac{\hat{l}^{2}}{p^{2}}|\psi_{njm}\rangle,
\end{eqnarray}
where $\hat{l}^{2}$ is the orbital angular momentum operator. Using
(\ref{eq:RadialAnsatz}) and the asymptotic expression for the radial
wavefunction, this energy shift is 
\begin{eqnarray}
\delta E_{njm} & = & \frac{1}{4}\left[\left(j+\frac{3}{2}\right)\left(j+\frac{1}{2}\right)+\left(j+\frac{1}{2}\right)\left(j-\frac{1}{2}\right)\right]\times\nonumber \\
 &  & \frac{1}{2^{n}n!\sqrt{\pi}}\int_{0}^{\infty}dp\,\frac{\left(H_{n}(p-v)\right)^{2}e^{-(p-v)^{2}}}{p^{2}}.\label{eq:Divergent}
\end{eqnarray}
Formally, the integral in (\ref{eq:Divergent}) is divergent as $p\rightarrow0$.
However, we can cure this by introducing a low energy cutoff, and
subtracting the divergent contribution. The integral is then dominated
near $p=v$, and can be well approximated by $\int_{0}^{\infty}dp\,\frac{\left(H_{n}(p-v)\right)^{2}e^{-(p-v)^{2}}}{p^{2}}=\frac{2^{n}n!\sqrt{\pi}}{v^{2}}+\mathcal{O}\left(\frac{1}{v^{3}}\right)$,
which is valid as long as $\left(H_{n}(p-v)\right)^{2}e^{-(p-v)^{2}}$
is localized in a region away from $p=0$. The lowest order shift
in energies is $\delta E_{njm}=\frac{\left(j+\frac{1}{2}\right)^{2}}{2v^{2}}$,
and the spectrum in the asymptotic limit is 
\begin{equation}
E_{njm}=\left(n+\frac{1}{2}\right)+\frac{\left(j+\frac{1}{2}\right)^{2}}{2v^{2}}-\frac{v^{2}}{2},\label{eq:ApproxSpectrum}
\end{equation}
consistent with the spectrum found using numerical diagonalization
in the previous section.

\subsubsection{Validity of Approximation}

To derive (\ref{eq:ApproxSpectrum}), we considered our particle in
a combination of two ``spin''-dependent potentials. The first is
the centrifugal barrier, and the second is the spin-orbit coupling.
The centrifugal barrier is repulsive, and divergent at $p=0$. The
spin-orbit term will form a well centered at $p=v.$ For sufficiently
large barriers, the minimum of the well will be far from the region
where the centrifugal barrier is finite. This implies that low energy
states will be well localized in the potential minimum produced by
the spin-orbit coupling. Since the well is approximately harmonic
near the minimum, the radial wavefunctions will be given by the one-dimensional
harmonic oscillator. These states are exponentially localized near
$p=v$, and will have minimum overlap with the centrifugal potential. 

Wavefunctions with larger radial quantum numbers will be increasingly
delocalized. The centrifugal barrier cannot be neglected when the
momentum-space wavefunction is finite near $p=0$. In this limit,
the two states with $j=l\pm s$ become mixed with the centrifugal
barrier. For higher radial quantum numbers, the system is better described
by two radial harmonic oscillators with angular momentum $l=j\pm s$,
and the spin-orbit term acts to mix the two states. Thus, even when
$v\gg1$, there will be a critical atomic number for which higher
energy states have average momentum $\langle p\rangle\sim2v$, where
the $p^{2}/2$ kinetic term dominates the $-vp$ of the spin-orbit
coupling. Above this threshold, the effect of the spin-orbit coupling
will become a perturbation, and the spectrum will approximate the
three-dimensional harmonic oscillator.

\section{Ground State}

In the previous section we showed the momentum space ground state
of a trapped particle with Weyl coupling has the approximate form
\begin{equation}
\psi_{0,\frac{1}{2},\pm\frac{1}{2}}(\mathbf{p})=\frac{1}{\sqrt{2}}\left(\chi_{\frac{1}{2},\pm\frac{1}{2}}^{-}(\hat{\mathbf{p}})-\chi_{\frac{1}{2},\pm\frac{1}{2}}^{+}(\hat{\mathbf{p}})\right)\frac{e^{-\left(p-v\right)^{2}/2}}{\pi^{1/4}}
\end{equation}
for $v\gg1$. The position space wavefunction can be found using the
radial differential equations. Alternatively, we can directly apply
the Fourier transform to the momentum-space wavefunction, $\tilde{\psi}_{0}(\mathbf{r})=\int d^{3}\mathbf{p}e^{i\mathbf{p}\cdot\mathbf{r}}\psi_{0}(\mathbf{p})$.
This is most easily evaluated by expanding the exponent $e^{i\mathbf{p}\cdot\mathbf{r}}=4\pi\sum_{lm}i^{l}j_{l}(pr)Y_{l}^{m}(\hat{\mathbf{r}})\left(Y_{l}^{m}(\hat{\mathbf{p}})\right)^{*}$.
Integration over the angular coordinate converts the spinor $\chi_{jm}^{\pm}(\hat{\mathbf{p}})$
to $\chi_{jm}^{\pm}(\hat{\mathbf{r}})$. The radial component is then
found from the integral $\int_{0}^{\infty}p^{2}j_{\frac{1}{2}\pm\frac{1}{2}}(pr)\frac{e^{-(p-v)^{2}/2}}{\pi^{1/4}}dp.$
The exponential factor localizes the integrand to a region near $p\sim v$,
in the asmyptotic regime with $v\gg1$, this integral can be evaluated
by extending the lower limit to $-\infty$, and then using the explicit
form of the spherical Bessel functions, $j_{0}(x)=\frac{\sin x}{x}$
and $j_{1}(x)=\frac{\cos x}{x}-\frac{\sin x}{x^{2}}$. \cite{DLMF_SphBess}
The position-space radial wavefunctions for the ground state are therefore
given by
\begin{eqnarray}
f_{0}(r) & = & \frac{\sqrt{2}}{\pi^{3/4}}\frac{1}{r}\left(r\cos(rv)+v\sin(rv)\right)e^{-r^{2}/2}\\
f_{1}(r) & = & \frac{\sqrt{2}}{\pi^{3/4}}\frac{1}{r^{2}}\left((1+r^{2})\sin(rv)-rv\cos(rv)\right)e^{-r^{2}/2}
\end{eqnarray}
with asymptotic corrections of $\mathcal{O}(1/v^{2})$. The full position
space wavefunction is approximately given by 
\begin{equation}
\psi_{0,\frac{1}{2},\pm\frac{1}{2}}(\mathbf{r})=\frac{1}{\sqrt{2}}\left(f_{0}(r)\chi_{\frac{1}{2},\pm\frac{1}{2}}^{-}(\hat{\mathbf{r}})-if_{1}(r)\chi_{\frac{1}{2},\pm\frac{1}{2}}^{+}(\hat{\mathbf{r}})\right)
\end{equation}
 in the asymptotic limit.

\subsection{Spin Textures and Currents}

\begin{figure}
(a) Spin current \hspace{.15\columnwidth}(b) Orbital current

\includegraphics[width=0.45\columnwidth]{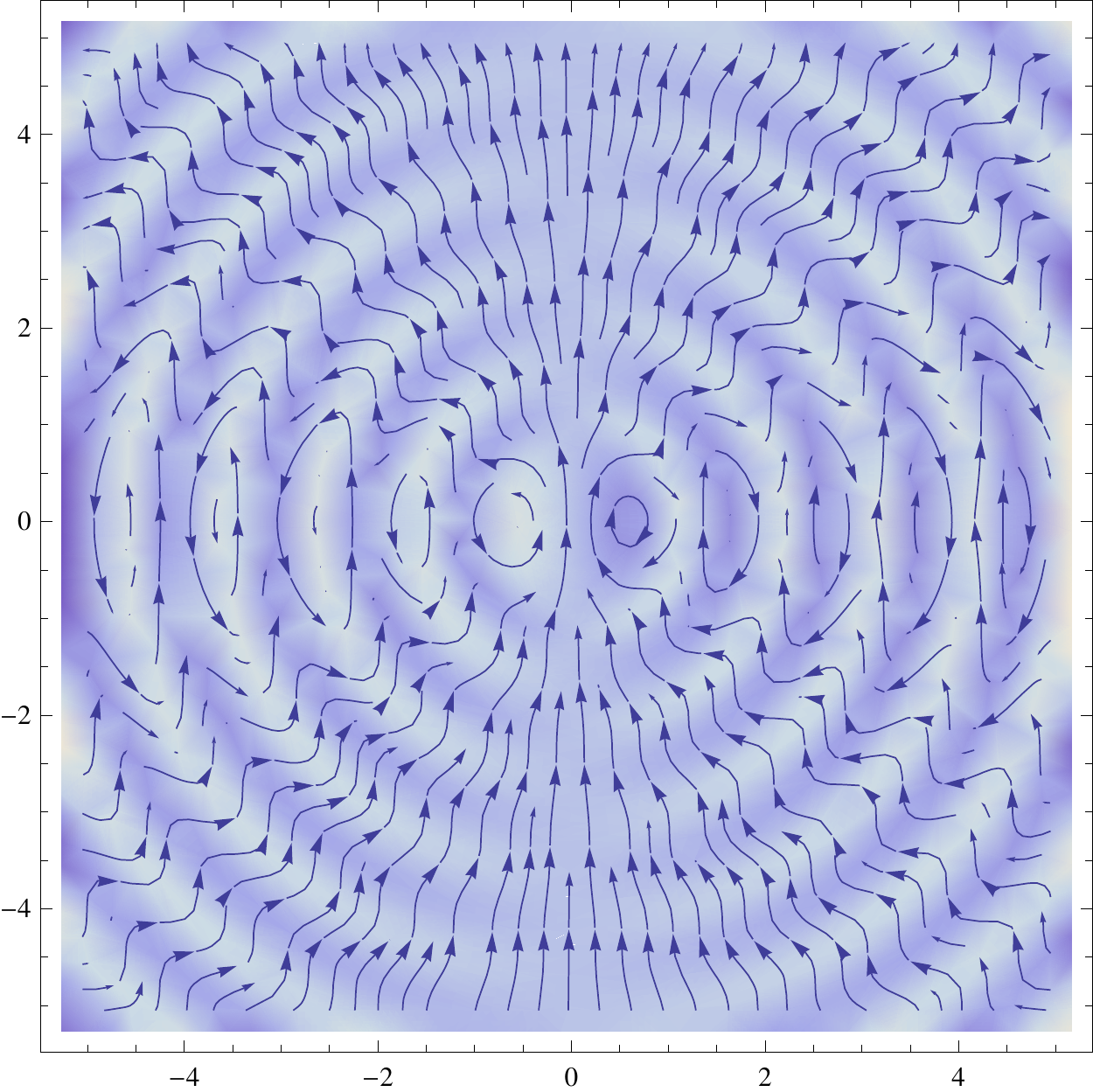}\includegraphics[width=0.45\columnwidth]{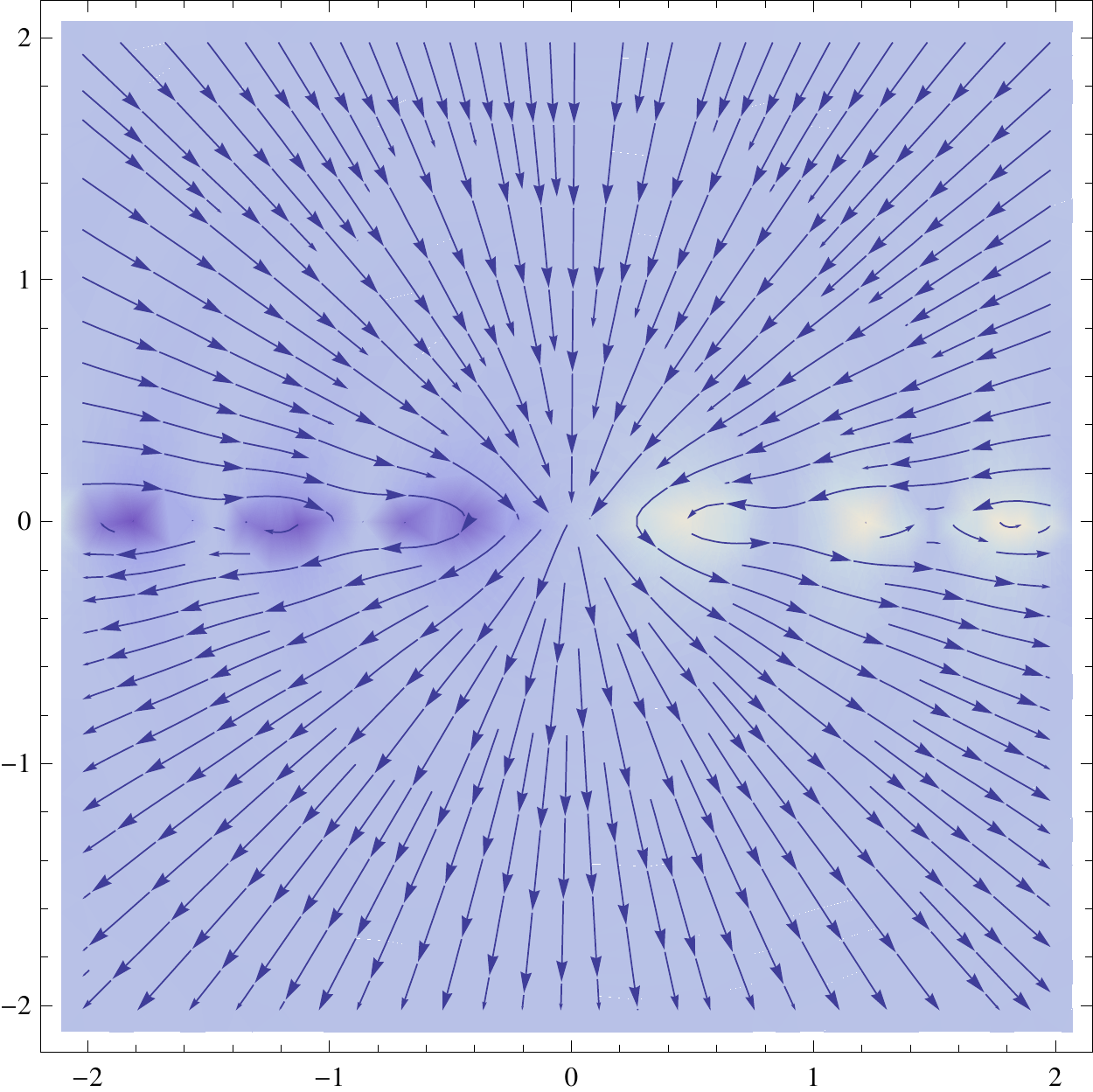}

(c) Total current

\includegraphics[width=0.45\columnwidth]{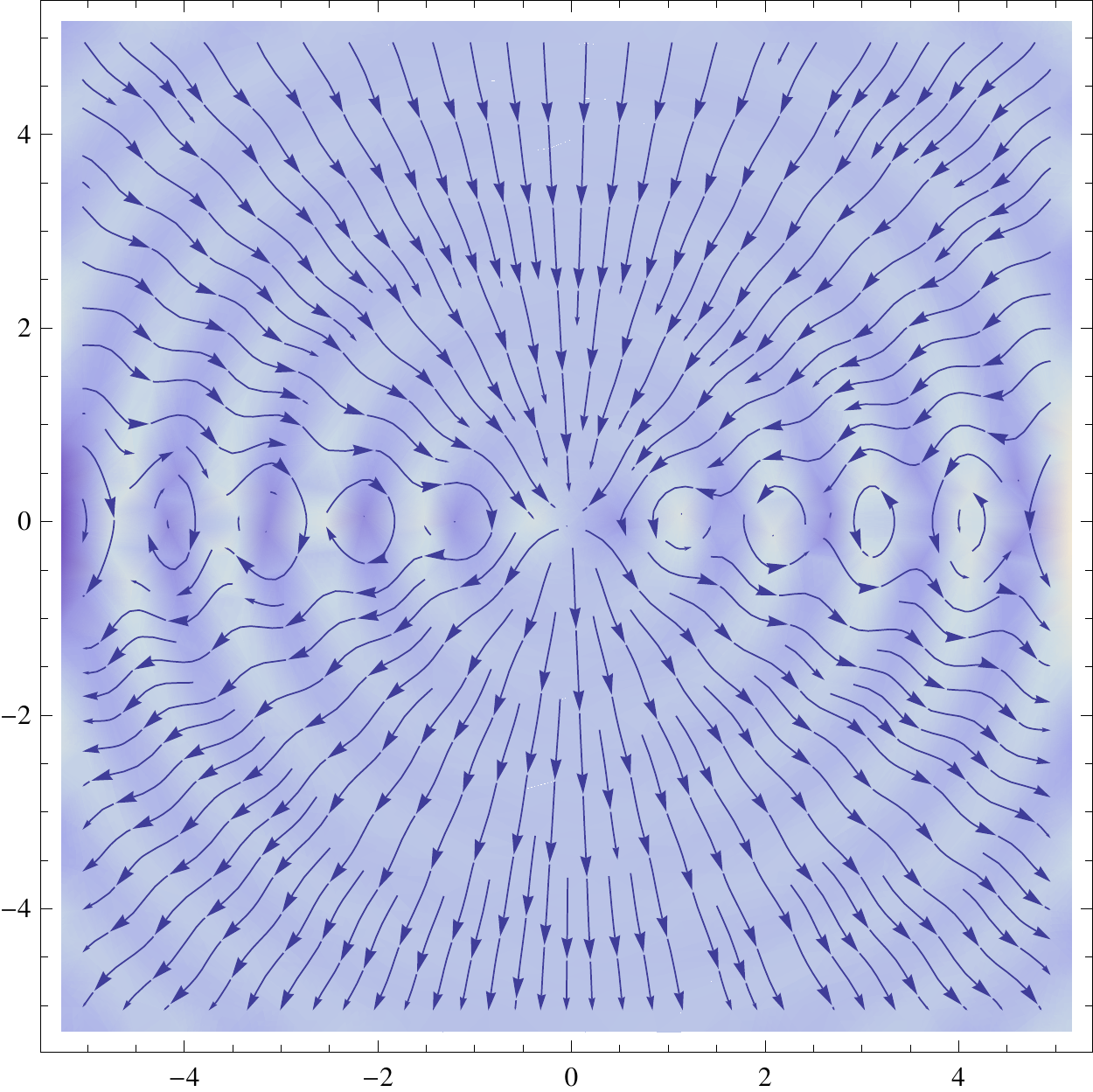}\caption{Spin, orbital and total currents for the ground state of a trapped
particle with Weyl coupling, with $j=\frac{1}{2}$, $m=\frac{1}{2}$
and $v=5$. In all figures the plane defined by $y=0$ plotted. The
arrows represent flows of the local normalized current vector. The
color image represents the out of plane component of the spin textures.
All three currents are azimuthally symmetric (a) The spin currents
have oscillations on the length scale $r\sim1/v$. On the axis with
$z=0$, the local spin vector is polarized entirely out of plane at
solutions to $\tan rv=-rv/v^{2}$. The odd solutions feature in-plane
vortex loops of spin, while the spin forms anti-vortices at the even
solutions. (b) The orbitals are dominated by the in-plane component,
which is stronger than the out of plane components by a factor of
$v$. A small out of plane component is largest on the $z=0$ axis.
In the upper half plane with $y>0$, current converges on the point
$r=0$, while on the lower half plane all the current diverges away
from the point $r=0$. (c) The total current is the sum of orbital
and spin currents. The total, spin and orbital currents are all independently
conserved.}
\label{fig:Currents}
\end{figure}

\subsubsection{Momentum Space}

The azimuthal component of the spinor $\chi_{\frac{1}{2}\frac{1}{2}}^{+}$
suggests the Weyl coupling should have a non-zero current in the ground
state. We first define the current operator in momentum space in the
standard way. First we multiply Schr\"{o}dinger's equation by $\psi^{\dagger}$,
$-i\psi^{\dagger}\frac{\partial}{\partial t}\psi=\psi^{\dagger}\left[-\frac{\nabla_{\mathbf{p}}^{2}}{2}+\frac{\mathbf{p}^{2}}{2}\right]\psi+v\mathbf{p}\cdot\left(\psi^{\dagger}\boldsymbol{\sigma}\psi\right)$,
and subtract the result from it's complex conjugate. The momentum-space
continuity equation is 
\begin{eqnarray}
\partial_{t}\left(\psi^{\dagger}\psi\right) & = & \nabla_{\mathbf{p}}\cdot\mathbf{j}_{\mathbf{p}}
\end{eqnarray}
where
\begin{equation}
\mathbf{j}_{\mathbf{p}}=\left(-\frac{i}{2}\right)\left[\psi^{\dagger}\nabla_{\mathbf{p}}\psi-\psi^{T}\nabla_{\mathbf{p}}\psi^{*}\right]=\Im\left[\psi^{\dagger}\nabla_{\mathbf{p}}\psi\right]
\end{equation}
is the momentum-space current density. We find that the ground state
current in momentum space is given by
\begin{equation}
\mathbf{j}_{\mathbf{p}}=\pm\frac{\sin\theta}{p}\frac{e^{-(p-v)^{2}}}{4\pi}\hat{\phi}
\end{equation}
for a state with $m=\pm\frac{1}{2}$.

\subsubsection{Position Space}

In position space, the spin-orbit coupling is imaginary, so the continuity
equation will have contributions from the orbital and spin degrees
of freedom: 
\begin{eqnarray}
\mathbf{j} & = & \mathbf{j}_{0}+\mathbf{j}_{s}
\end{eqnarray}
where the orbital current $\mathbf{j}_{0}=\Im\left[\psi^{\dagger}\nabla\psi\right]$
is analogous to the orbital current in momentum space space, and the
spin current is $\mathbf{j}_{s}=v\left(\psi^{\dagger}\boldsymbol{\mathbf{\sigma}}\sigma\right)$.
For an eigenstate of the trapped Weyl Hamiltonian, these currents
are azimuthally symmetric. The ground state currents can be calculated
exactly for large spin-orbit coupling, and are given by\begin{widetext}
\begin{eqnarray}
\mathbf{j}_{0} & = & \frac{1}{4\pi}\left[\left(f_{0}^{+}\partial_{r}f_{0}^{-}-f_{0}^{-}\partial_{r}f_{0}^{+}\right)\cos\theta\hat{r}+\sin\theta\frac{f_{0}^{+}}{r}\left(f_{0}^{-}\hat{\theta}+f_{0}^{+}\hat{\phi}\right)\right]\\
\mathbf{j}_{s} & = & \frac{v}{4\pi}\left[\left(\left(f_{0}^{+}\right)^{2}\cos\theta-\left(f_{0}^{-}\right)^{2}\sin\theta\right)\hat{r}+\left(\left(f_{0}^{+}\right)^{2}\sin\theta+\left(f_{0}^{-}\right)^{2}\cos\theta\right)\hat{\theta}+\left(2f_{0}^{+}f_{0}^{-}\sin\theta\right)\hat{\phi}\right]
\end{eqnarray}
\end{widetext}for the ground state with $j=\frac{1}{2}$ and $m=\frac{1}{2}$.
These currents are shown in Fig.\ref{fig:Currents}. For a plane defined
by fixed $\phi$, both currents are azimuthally invariant: that is
the current fields do not change forms under a rotation by $\phi$.
Since the length scale $1/v$ appears only in the radial variable,
the radial component of the orbital current will be stronger than
the polar and azimuthal component by a factor of $v$. As is seen
in Fig. \ref{fig:Currents}(b), the out of plane component of spin
is negligible only near the plane defined by $\theta=\pi/2$. The
lower half-plane has all currents flowing to the point $r=0$, while
all currents in the upper half plane diverge from this point. 

The spin currents do have an out of plane, or azimuthal, component
of spin. The in-plane component has a similar structure to the orbital
current, in the lower half-plane the current converges to $r=0$,
while it diverges from the point $r=0$ in the upper half plane. The
out-of-plane component oscillates on the scale $r\sim1/v$ with amplitude
$\cos\theta$. At special points corresponding to the solutions of
$\tan rv=-rv/v^{2}$, the out of plane spin current is completely
polarized in the azimuthal direction. The sum of the two currents
is seen in Fig. \ref{fig:Currents}(c). Since the orbital and spin
currents are independently conserved, the total current is also conserved.

\section{Conclusion}

The problem of the three-dimensional spin-orbit coupling in a harmonic
trap was considered. The system has a conserved total angular momentum
that allows the Hamiltonian to be projected into sectors where $j,m$
are good quantum numbers. Each sector is tri-diagonal, providing for
an efficient numerical calculation of the spectrum. At large values
of the spin-orbit coupling parameter, the system undergoes a dimensional
reduction from three to one. The spectrum is well approximated by
Landau levels in the radial coordinate, with splittings inversely
proportional the the spin-orbit coupling strength. 

The conservation of total angular momentum allows us to express Schr\"{o}dinger's
equation as a set of coupled differential equations in the radial
coordinate. We find the form of these differential equations in both
position and momentum space. Using asymptotic analysis on the momentum
space differential equations, we reproduce an analytic spectrum that
well approximates the spectrum found numerically. We further use this
approximation to find analytical expressions for the low energy eigenfunctions
in both position and momentum space. 

Finally, we explore the properties of the ground state wavefunction.
We find a momentum space orbital current in the azimuthal direction.
In position space the total current can be decomposed into a spin
current and an orbital current. All currents are invariant under azimuthal
rotation. At large spin-orbit coupling, the spin currents have an
azimuthal component that oscillates on the inverse of the spin-orbit
strength. Along the $z=0$ axis, the spin currents alternate between
complete polarization with and against the azimuthal plane. At these
special points, the in plane spin density has vortex and anti-vortex
structure. The orbital current is characterized by a convergence of
all currents at the point $r=0$ in the upper half-sphere, with a
divergence of away from the same point in the lower half-sphere. 

A single trapped particle with Weyl coupling has rich ground state
textures. This suggests that a system of trapped bosonic atoms with
Weyl coupling will have novel many-body phases. Early evidence shows
such a system will have qualitatively new ground states, such as a
cubic lattice phase. \cite{3Dbosons,Scooped.Wu,Scooped.Machida}

\section{Acknowledgements}

During the completion of this manuscript, the authors became aware
of two similar works that reproduce the skyrmion spin textures.\cite{Scooped.Machida,Scooped.Wu}

This research was performed in part under the sponsorship of the US
Department of Commerce, National Institute of Standards and Technology,
and was supported by the National Science Foundation under Physics
Frontiers Center Grant PHY-0822671 and by the ARO under the DARPA
OLE program.

\bibliographystyle{apsrev4-1.bst}
\bibliography{3DSOC}

\section{Appendix A\label{sec:Appendix-A}}

\begin{widetext}In this appendix we show that $\Op$ and $\Om$ have
the matrix elements as described by Eq.\ref{eq:Aopm} and \ref{eq:Aopp}.
To do this, we first need to calculate the matrix elements of the
three-dimensional isotropic harmonic oscillator basis, $\langle n^{\prime}l^{\prime}m^{\prime}|a_{q}|nlm\rangle$
and $\langle n^{\prime}l^{\prime}m^{\prime}|a_{q}^{\dagger}|nlm\rangle$,
where $q=0,\pm1$, and the states $|nlm\rangle$ are the states three-dimensional
harmonic oscillator basis as defined in the main text.

It is convenient to define the number basis $\ket{n,l,m_{l}}$ where
$(n,l,m_{l})$ are respectively the radial, angular and magnetic quantum
numbers. In the absence of spin-orbit coupling we have the harmonic
oscillator energies $E=\hbar\omega\left(2n+l+\frac{3}{2}\right)$
where a state of angular momentum $l$ has a $2l+1$-fold degeneracy
of $m_{l}$. To do this we define the spherical creation and annihilation
operators $a_{\pm}=\mp\frac{1}{\sqrt{2}}\left(a_{x}\mp ia_{y}\right)$,
$a_{\pm}^{\dagger}=\mp\frac{1}{\sqrt{2}}\left(a_{x}^{\dagger}\pm ia_{y}^{\dagger}\right)$,
and $a_{0}=a_{z}$, $a_{0}^{\dagger}=a_{z}^{\dagger}$. We will find
it helpful to express the angular momentum generators as 
\begin{eqnarray}
L_{+} & = & \sqrt{2}\left(a_{+}^{\dagger}a_{z}+a_{z}^{\dagger}a_{-}\right),\\
L_{-} & = & \sqrt{2}\left(a_{-}^{\dagger}a_{z}+a_{z}^{\dagger}a_{+}\right),\\
L_{z} & = & a_{+}^{\dagger}a_{+}-a_{-}^{\dagger}a_{-}
\end{eqnarray}
with $L_{x}=\frac{1}{2}\left(L_{+}+L_{-}\right)$ and $L_{y}=-\frac{i}{2}\left(L_{+}-L_{-}\right)$.
If we additionally define the operators 
\begin{eqnarray}
S_{+} & = & \frac{1}{2}\left(a_{0}^{\dagger}\right)^{2}-a_{+}^{\dagger}a_{-}^{\dagger}\\
S_{-} & = & \frac{1}{2}\left(a_{0}\right)^{2}-a_{+}a_{-}\\
S_{0} & = & \frac{1}{2}\left(\hat{N}+\frac{3}{2}\right)
\end{eqnarray}
These operators commute with the angular momentum operators, and satisfy
the commutation relations $[S_{+},S_{-}]=-2S_{0}$ and $[S_{0},S_{\pm}]=\pm S_{\pm}$.
The three-dimensional harmonic oscillator eigenstates can be expressed
as 
\begin{equation}
|nlm\rangle=\left[\frac{\Gamma\left(l+\frac{3}{2}\right)}{n!\Gamma\left(n+l+\frac{3}{2}\right)}\frac{(l+m)!}{(2l)!(l-m)!}\right]S_{+}^{n}L_{-}^{l-m}|0,l,l\rangle\label{eq:nlm1-1}
\end{equation}
where $|0,l,l\rangle=\frac{1}{\sqrt{l!}}\left(a_{+}^{\dagger}\right)^{l}|0,0,0\rangle$
is a state of maximum angular momentum. It can be seen that a state
$|nlm\rangle$ has energy $E=2n+l+\frac{3}{2}$.

\subsection{Caculation of: $a_{q}^{\dagger}$}

\begin{lyxgreyedout}
\global\long\def\np{n^{\prime}}
\global\long\def\mpp{m^{\prime}}
\global\long\def\lp{l^{\prime}}
\global\long\def\Nt{\tilde{\mathcal{N}}}
\global\long\def\N{\mathcal{N}}
\end{lyxgreyedout}
To calculate $a_{-}^{\dagger}$, we first note that the operator changes
the total energy number by $\Delta n=+1$ unit, so it must connect
states with either $\Delta n=+1$ and $\Delta l=-1$, or $\Delta n=0$
and $\Delta l=+1$. Other states with $\Delta n>1$ and $\Delta l=-1+2(\Delta n-1)$
are consistent with this condition, but must have zero matrix elements
from the fact that $a_{q}$ is a spherical tensor of rank-1, and higher
order corrections are inconsistent with angular momentum conservation.
Using this fact, the matrix elements can be decomposed as

\begin{eqnarray}
\bra{\np,\lp,\mpp}a_{-}^{\dagger}\ket{n,l,m} & = & \bra{n,l+1,m-1}a_{-}^{\dagger}\ket{n,l,m}\delta_{\np,n}\delta_{\lp,l+1}\delta_{\mpp,m-1}\\
 & + & \bra{n+1,l-1,m-1}a_{-}^{\dagger}\ket{n,l,m}\delta_{\np,n+1}\delta_{\lp,l-1}\delta_{\mpp,m-1}
\end{eqnarray}
The two non-zero elements can be calculated individually. We first
calculate $\bra{n,l+1,m-1}a_{-}^{\dagger}\ket{n,l,m}$. It is convenient
to use the equivalent definition of $\ket{n,l,m}$, given by 

\begin{equation}
|nlm\rangle=\left[\frac{\Gamma\left(l+\frac{3}{2}\right)}{n!\Gamma\left(n+l+\frac{3}{2}\right)}\frac{(l-m)!}{(2l)!(l+m)!}\right]S_{+}^{n}L_{-}^{l+m}|0,l,-l\rangle
\end{equation}
to express the desired element as 
\begin{eqnarray}
\bra{n,l+1,m-1}a_{-}^{\dagger}\ket{n,l,m} & = & \Nt_{n,l+1,m-1}\Nt_{n,l,m}\bra{{\bf 0}}(a_{-})^{l+1}(L_{-})^{l+m}(S_{-})^{n}a_{-}^{\dagger}(S_{+})^{n}(L_{+})^{l+m}(a_{-}^{\dagger})^{l}\ket{{\bf 0}}\\
 & = & \Nt_{n,l+1,m-1}\Nt_{n,l,m}\bra{{\bf 0}}(a_{-})^{l+1}(L_{-})^{l+m}(S_{-})^{n}(S_{+})^{n}a_{-}^{\dagger}(L_{+})^{l+m}(a_{-}^{\dagger})^{l}\ket{{\bf 0}}
\end{eqnarray}
where the normalization constant is $\Nt_{n,l,m}=\left[\frac{\Gamma\left(l+\frac{3}{2}\right)}{n!\Gamma\left(n+l+\frac{3}{2}\right)}\frac{(l-m)!}{(2l)!(l+m)!}\right]$,
and we have used the commutation relation $[a_{-}^{\dagger},S_{+}]=0$.
The bra $\bra 0(a_{-})^{l+1}(L_{-})^{l+m}$ is an eigenbra of the
the operator $\left(S_{-}\right)^{n}\left(S_{+}\right)^{n}$ with
eigenvalue $\frac{n!\Gamma(n+l+3/2+1)}{\Gamma(l+3/2+1)}$. We can
again use the commutation of the operator $[L_{-},a_{-}^{\dagger}]=0$
to find the remaining factor 
\begin{eqnarray}
\bra{{\bf 0}}(a_{-})^{l+1}(L_{-})^{l+m}a_{-}^{\dagger}(L_{+})^{l+m}(a_{-}^{\dagger})^{l}\ket{{\bf 0}} & = & \bra{{\bf 0}}(a_{-})^{l+1}a_{-}^{\dagger}(L_{-})^{l+m}(L_{+})^{l+m}(a_{-}^{\dagger})^{l}\ket{{\bf 0}}\\
 & = & (l+1)\bra{{\bf 0}}(a_{-})^{l}(L_{-})^{l+m}(L_{+})^{l+m}(a_{-}^{\dagger})^{l}\ket{{\bf 0}}\\
 & = & (l+1)\Nt_{0,l,m}^{-2}.
\end{eqnarray}
These factors combine to give the non-zero value of the matrix element
\begin{equation}
\bra{n,l+1,m-1}a_{-}^{\dagger}\ket{n,l,m}=\frac{1}{2}\sqrt{\frac{n+l+3/2}{l+3/2}\frac{(l-m+2)(l-m+1)}{l+1/2}}.
\end{equation}
Similar techniques can applied to evaluate the other non-zero matrix
element of the operator $a_{-}^{\dagger}$. The calculation is straightforward
using the same techniques. The full matrix element is given by

\begin{eqnarray}
\bra{\np,\lp,\mpp}a_{-}^{\dagger}\ket{n,l,m} & = & \frac{1}{2}\sqrt{\frac{n+l+3/2}{l+3/2}\frac{(l-m+2)(l-m+1)}{l+1/2}}\delta_{\np,n}\delta_{\lp,l+1}\delta_{\mpp,m-1}\\
 & - & \frac{1}{2}\sqrt{\frac{n+1}{l+1/2}\frac{(l+m)(l+m-1)}{l-1/2}}\delta_{\np,n+1}\delta_{\lp,l-1}\delta_{\mpp,m-1}.
\end{eqnarray}

The remaining matrix elements can be calculated in an analogous way.
The results can be summarized by the expression

\begin{eqnarray}
a_{q}^{\dagger}\ket{n,l,m} & = & c_{q}^{+}(n,l,m)\ket{n,l+1,m+q}+d_{q}^{+}(n,l,m)\ket{n+1,l-1,m+q}
\end{eqnarray}
 where $q=-1,0,1$, and the matrix elements are given by 
\begin{eqnarray}
c_{q}^{-}(n,l,m) & = & \left(\frac{1}{\sqrt{2}}\right)^{1+|q|}\sqrt{\frac{n+l+1/2}{(l+1/2)(l-1/2)}}f_{q}(l,m)\\
d_{q}^{-}(n,l,m) & = & (-1)^{q}\left(\frac{1}{\sqrt{2}}\right)^{1+|q|}\sqrt{\frac{n}{(l+3/2)(l+1/2)}}g_{-q}(l,m)\\
c_{q}^{+}(n,l,m) & = & \left(\frac{1}{\sqrt{2}}\right)^{1+|q|}\sqrt{\frac{n+l+3/2}{(l+3/2)(l+1/2)}}g_{q}(l,m)\\
d_{q}^{+}(n,l,m) & = & (-1)^{q}\left(\frac{1}{\sqrt{2}}\right)^{1+|q|}\sqrt{\frac{n+1}{(l+1/2)(l-1/2)}}f_{-q}(l,m)
\end{eqnarray}
 where the functions $f_{q}(l,m)$ and $g_{q}(l,m)$ are defined as
\begin{eqnarray}
f_{q}(l,m) & = & \left\{ \begin{array}{ll}
\sqrt{(l+m)(l+m-1)} & q=+1\\
\sqrt{(l+m)(l-m)} & q=0\\
\sqrt{(l-m)(l-m-1)} & q=-1
\end{array}\right.\\
g_{q}(l,m) & = & \left\{ \begin{array}{ll}
\sqrt{(l+m+2)(l+m+1)} & q=+1\\
\sqrt{(l+m+1)(l-m+1)} & q=0\\
\sqrt{(l-m+2)(l-m+1)} & q=-1
\end{array}\right..
\end{eqnarray}
The matrix elements of the operators $a_{q}$ can be calculated through
conjugation.

\subsection{Matrix elements of $\Op$ and $\Om$.}

We can now calculate the matrix elements of $\Op$ and $\Om$ for
states of good total angular momentum labeled by quantum numbers $j,m$.
Recall in the main text that these states are defined by 
\begin{equation}
\ket{n,j,m,\lambda}=e^{i\phi_{\lambda}}\begin{pmatrix}\lambda\sqrt{\frac{j-\lambda m+x_{\lambda}}{2(j+x_{\lambda})}}\ket{n,j+\atwo,m-\onehalf}\\
-\sqrt{\frac{j+\lambda m+x_{\lambda}}{2(j+x_{\lambda})}}\ket{n,j+\atwo,m+\onehalf}
\end{pmatrix}
\end{equation}
where $\phi_{\lambda}$ is an arbitrary phase. We will find it convenient
to express this as 
\begin{equation}
\ket{n,j,m,\lambda}=\begin{pmatrix}\gamma_{\lambda}^{\uparrow}\ket{n,j+\atwo,m-\onehalf}\\
\gamma_{\lambda}^{\downarrow}\ket{n,j+\atwo,m+\onehalf}
\end{pmatrix},
\end{equation}
where $\gamma_{\lambda}^{\uparrow}=\lambda e^{i\phi_{\lambda}}\sqrt{\frac{j-\lambda m+x_{\lambda}}{2(j+x_{\lambda})}}$
and $\gamma_{\lambda}^{\downarrow}=-e^{i\phi_{\lambda}}\sqrt{\frac{j+\lambda m+x_{\lambda}}{2(j+x_{\lambda})}}$.

We now consider the action of the operator $\boldsymbol{\sigma}\cdot\mathbf{p}$
on the states $\ket{n,j,m,\lambda}$,

\begin{eqnarray}
\left(\sqrt{2}\left(\sigma_{-}a_{-}-\sigma_{+}a_{+}\right)+\sigma_{z}a_{z}\right)\ket{n,j,m,\lambda} & = & \begin{pmatrix}-\sqrt{2}\gamma_{\lambda}^{\downarrow}a_{+}\ket{n,j+\atwo,m+\onehalf}+\gamma_{\lambda}^{\uparrow}a_{z}\ket{n,j+\atwo,m-\onehalf}\\
\sqrt{2}\gamma_{\lambda}^{\uparrow}a_{-}\ket{n,j+\atwo,m-\onehalf}-\gamma_{\lambda}^{\downarrow}a_{z}\ket{n,j+\atwo,m+\onehalf}
\end{pmatrix}\\
 & = & \begin{pmatrix}\left(-\sqrt{2}\gamma_{\lambda}^{\downarrow}c_{+}^{-}(n,j+\atwo,m+\onehalf)+\gamma_{\lambda}^{\uparrow}c_{0}^{-}(n,j+\atwo,m-\onehalf)\right)\ket{n,j-1+\atwo,m-\onehalf}\\
\left(\sqrt{2}\gamma_{\lambda}^{\uparrow}c_{-}^{-}(n,j+\atwo,m-\onehalf)-\gamma_{\lambda}^{\downarrow}c_{0}^{-}(n,j+\atwo,m+\onehalf)\right)\ket{n,j-1+\atwo,m+\onehalf}
\end{pmatrix}\\
 & + & \begin{pmatrix}\left(-\sqrt{2}\gamma_{\lambda}^{\downarrow}d_{+}^{-}(n,j+\atwo,m+\onehalf)+\gamma_{\lambda}^{\uparrow}d_{0}^{-}(n,j+\atwo,m-\onehalf)\right)\ket{n-1,j+1+\atwo,m-\onehalf}\\
\left(\sqrt{2}\gamma_{\lambda}^{\uparrow}d_{-}^{-}(n,j+\atwo,m-\onehalf)-\gamma_{\lambda}^{\downarrow}d_{0}^{-}(n,j+\atwo,m+\onehalf)\right)\ket{n-1,j+1+\atwo,m+\onehalf}
\end{pmatrix}.
\end{eqnarray}
This can be expressed as
\begin{eqnarray}
\left(\sqrt{2}\left(\sigma_{-}a_{-}-\sigma_{+}a_{+}\right)+\sigma_{z}a_{z}\right)\ket{n,j,m,\lambda} & = & \begin{pmatrix}\Gamma_{c}^{\uparrow}(\lambda)\ket{n,j-1+\atwo,m-\onehalf}\\
\Gamma_{c}^{\downarrow}(\lambda)\ket{n,j-1+\atwo,m+\onehalf}
\end{pmatrix}+\begin{pmatrix}\Gamma_{d}^{\uparrow}(\lambda)\ket{n-1,j+1+\atwo,m-\onehalf}\\
\Gamma_{d}^{\downarrow}(\lambda)\ket{n-1,j+1+\atwo,m+\onehalf}
\end{pmatrix},
\end{eqnarray}
where the coefficients 
\begin{eqnarray}
\Gamma_{c}^{\uparrow}(\lambda) & = & \left(-\sqrt{2}\gamma_{\lambda}^{\downarrow}c_{+}^{-}\left(n,j+\atwo,m+\onehalf\right)+\gamma_{\lambda}^{\uparrow}c_{0}^{-}\left(n,j+\atwo,m-\onehalf\right)\right)\\
 & = & -e^{i\phi_{\lambda}}\sqrt{\frac{n+j+\atwo+\onehalf}{2(j+\atwo+\onehalf)(j+\atwo-\onehalf)}}\left(-f_{+}\left(j+\atwo,m+\onehalf\right)\gamma_{\lambda}^{\downarrow}+f_{0}\left(j+\atwo,m-\onehalf\right)\gamma_{\lambda}^{\uparrow}\right)\\
 & = & e^{i\phi_{\lambda}}\sqrt{\frac{\left(n+j+x_{\lambda}\right)\left(j+m+x_{\lambda}-1\right)}{4(j+x_{\lambda})^{2}(j+x_{\lambda}-1)}}\left(\sqrt{\left(j+\lambda m+x_{\lambda}\right)\left(j+m+x_{\lambda}\right)}+\lambda\sqrt{\left(j-\lambda m+x_{\lambda}\right)\left(j-m+x_{\lambda}\right)}\right)
\end{eqnarray}
and
\begin{eqnarray}
\Gamma_{c}^{\downarrow}(\lambda) & = & \left(\sqrt{2}\gamma_{\lambda}^{\uparrow}c_{-}^{-}\left(n,j+\atwo,m-\onehalf\right)-\gamma_{\lambda}^{\downarrow}c_{0}^{-}\left(n,j+\atwo,m+\onehalf\right)\right)\\
 & = & -e^{i\phi_{\lambda}}\sqrt{\frac{n+j+\atwo+\onehalf}{2(j+\atwo+\onehalf)(j+\atwo-\onehalf)}}\left(-f_{-}\left(j+\atwo,m-\onehalf\right)\gamma_{\lambda}^{\uparrow}-f_{0}\left(j+\atwo,m+\onehalf\right)\gamma_{\lambda}^{\downarrow}\right)\\
 & = & e^{i\phi_{\lambda}}\sqrt{\frac{\left(n+j+x_{\lambda}\right)\left(j-m+x_{\lambda}-1\right)}{4(j+x_{\lambda})^{2}(j+x_{\lambda}-1)}}\left(\lambda\sqrt{\left(j-\lambda m+x_{\lambda}\right)\left(j-m+x_{\lambda}\right)}+\sqrt{\left(j+\lambda m+x_{\lambda}\right)\left(j+m+x_{\lambda}\right)}\right).
\end{eqnarray}
Explicitly calculating these for $\lambda=\pm1$ we get: 
\begin{eqnarray}
\left(\sqrt{2}\gamma_{+}^{\downarrow}c_{+}^{-}\left(n,j+\onehalf,m+\onehalf\right)+\gamma_{+}^{\uparrow}c_{0}^{-}\left(n,j+\onehalf,m-\onehalf\right)\right) & = & e^{i\phi_{+}}\sqrt{2(n+j+1)}\sqrt{\frac{j+m}{2j}}\\
\left(\sqrt{2}\gamma_{-}^{\downarrow}c_{+}^{-}\left(n,j-\onehalf,m+\onehalf\right)+\gamma_{-}^{\uparrow}c_{0}^{-}\left(n,j-\onehalf,m-\onehalf\right)\right) & = & 0
\end{eqnarray}
and 
\begin{eqnarray}
\left(\sqrt{2}\gamma_{+}^{\uparrow}c_{-}^{-}\left(n,j+\onehalf,m-\onehalf\right)-\gamma_{+}^{\downarrow}c_{0}^{-}\left(n,j+\onehalf,m+\onehalf\right)\right) & = & e^{i\phi_{+}}\sqrt{2(n+j+1)}\sqrt{\frac{j-m}{2j}}\\
\left(\sqrt{2}\gamma_{-}^{\uparrow}c_{-}^{-}\left(n,j-\onehalf,m-\onehalf\right)-\gamma_{-}^{\downarrow}c_{0}^{-}\left(n,j-\onehalf,m+\onehalf\right)\right) & = & 0.
\end{eqnarray}
 Similarly, we calculate the equivalent terms for $\Gamma_{d}^{s}(\lambda)$
with $s=\uparrow,\downarrow$. This gives 
\begin{eqnarray}
\Gamma_{d}^{\uparrow}(\lambda) & = & \left(-\sqrt{2}\gamma_{\lambda}^{\downarrow}d_{+}^{-}\left(n,j+\atwo,m+\onehalf\right)+\gamma_{\lambda}^{\uparrow}d_{0}^{-}\left(n,j+\atwo,m-\onehalf\right)\right)\\
 & = & e^{i\phi_{\lambda}}\sqrt{\frac{n\left(j-m+x_{\lambda}+1\right)}{4(j+x_{\lambda})^{2}(j+x_{\lambda}+1)}}\left(\sqrt{\left(j+\lambda m+x_{\lambda}\right)\left(j-m+x_{\lambda}\right)}-\lambda\sqrt{\left(j-\lambda m+x_{\lambda}\right)\left(j+m+x_{\lambda}\right)}\right)
\end{eqnarray}
\begin{eqnarray}
\Gamma_{d}^{\downarrow}(\lambda) & = & \left(\sqrt{2}\gamma_{\lambda}^{\uparrow}d_{-}^{-}\left(n,j+\atwo,m-\onehalf\right)-\gamma_{\lambda}^{\downarrow}d_{+}^{0}\left(n,j+\atwo,m+\onehalf\right)\right)\\
 & = & -e^{i\phi_{\lambda}}\sqrt{\frac{n\left(j+m+x_{\lambda}+1\right)}{4(j+x_{\lambda})^{2}(j+x_{\lambda}+1)}}\left(\lambda\sqrt{\left(j-\lambda m+x_{\lambda}\right)\left(j+m+x_{\lambda}\right)}-\sqrt{\left(j-m+x_{\lambda}\right)\left(j+\lambda m+x_{\lambda}\right)}\right)
\end{eqnarray}
Explicitly, 
\begin{eqnarray}
\Gamma_{d}^{\uparrow}(+)=\left(-\sqrt{2}\gamma_{+}^{\downarrow}d_{+}^{-}\left(n,j+\onehalf,m+\onehalf\right)+\gamma_{+}^{\uparrow}d_{0}^{-}\left(n,j+\onehalf,m-\onehalf\right)\right) & = & 0\\
\Gamma_{d}^{\uparrow}(-)=\left(-\sqrt{2}\gamma_{-}^{\downarrow}d_{+}^{-}\left(n,j-\onehalf,m+\onehalf\right)+\gamma_{-}^{\uparrow}d_{0}^{-}\left(n,j-\onehalf,m-\onehalf\right)\right) & = & e^{i\phi_{-}}\sqrt{2n}\sqrt{\frac{j-m+1}{2(j+1)}}
\end{eqnarray}
 
\begin{eqnarray}
\Gamma_{d}^{\downarrow}(+)=\left(\sqrt{2}\gamma_{+}^{\uparrow}d_{-}^{-}\left(n,j+\onehalf,m-\onehalf\right)-\gamma_{+}^{\downarrow}d_{+}^{0}\left(n,j+\onehalf,m+\onehalf\right)\right) & = & 0\\
\Gamma_{d}^{\downarrow}(-)=\left(\sqrt{2}\gamma_{-}^{\uparrow}d_{-}^{-}\left(n,j-\onehalf,m-\onehalf\right)-\gamma_{-}^{\downarrow}d_{+}^{0}\left(n,j-\onehalf,m+\onehalf\right)\right) & = & e^{i\phi_{-}}\sqrt{2n}\sqrt{\frac{j+m+1}{2(j+1)}}
\end{eqnarray}

We summarize these relations as (restoring the subscript on $n_{r}$)
\begin{eqnarray}
\Om\ket{n_{r},j,m,+} & = & e^{i\Delta\phi}\sqrt{2n_{r}}\ket{n_{r}-1,j,m,-}\\
\Om\ket{n_{r},j,m,-} & = & e^{-i\Delta\phi}\sqrt{2(n_{r}+j+1)}\ket{n_{r},j,m,+}
\end{eqnarray}
where $\Delta\phi=\left(\phi_{+}-\phi_{-}\right)$. We can find the
action of the operator $\Op={\Om}^{\dagger}$ through conjugation
\begin{eqnarray}
\Op\ket{n_{r},j,m,+} & = & e^{-i\Delta\phi}\sqrt{2(n_{r}+j+1)}\ket{n_{r},j,m,-}\\
\Op\ket{n_{r},j,m,-} & = & e^{i\Delta\phi}\sqrt{2(n_{r}+1)}\ket{n_{r}+1,j,m,+}
\end{eqnarray}
We can show that $\frac{1}{2}\{\Op,\Om\}=\hat{N}+3/2$, so 
\begin{equation}
H=\frac{1}{2}\{\Op,\Om\}+i\frac{v}{\sqrt{2}}(\Op-\Om).
\end{equation}
If we chose $e^{i\Delta\phi}=-1$, we see that the matrix elements
for ${\bf p}\cdot\boldsymbol{\sigma}=\frac{i}{\sqrt{2}}\left(\Op-\Om\right)$
are:
\begin{eqnarray}
\langle n_{r}^{\prime},j,m,+|\frac{i}{\sqrt{2}}\left(\Op-\Om\right)|n_{r},j,m,-\rangle & = & i\left(\sqrt{n_{r}+1}\delta_{n_{r}+1,n_{r}^{\prime}}-\sqrt{n_{r}+j+1}\delta_{n_{r},n_{r}^{\prime}}\right)\\
\langle n_{r}^{\prime},j,m,-|\frac{i}{\sqrt{2}}\left(\Op-\Om\right)|n_{r},j,m,+\rangle & = & i\left(\sqrt{n_{r}+j+1}\delta_{n_{r},n_{r}^{\prime}}-\sqrt{n_{r}}\delta_{n_{r}-1,n_{r}^{\prime}}\right).
\end{eqnarray}

\section{Appendix B\label{sec:Appendix-B}}

In this appendix we calculate the radial Schr\"{o}dinger equation
in position space by Fourier transforming the momentum-space version.
We want to find the Fourier transform of the eigenfunctions of the
harmonic oscillator. We first need to find the momentum space eigenfunctions
of the three-dimensional spherical harmonic oscillator int terms of
the momentum space harmonic oscillator wavefunctions$\langle\mathbf{p}|\psi_{nlm}\rangle=\int\frac{d\mathbf{r}}{\sqrt{(2\pi)^{3}}}e^{i\mathbf{p}\cdot\mathbf{r}}\langle\mathbf{r}|\psi_{nlm}\rangle$.
Using the expansion of plane waves into spherical harmonics,$e^{i\mathbf{p}\cdot\mathbf{r}}=4\pi\sum_{lm}i^{l}j_{l}(pr)Y_{l}^{m}(\hat{\mathbf{p}})\left(Y_{l}^{m}(\hat{\mathbf{r}})\right)^{*}$,
we find the relation 
\begin{eqnarray}
\langle\mathbf{p}|\psi_{nlm}\rangle & = & \int\frac{d\mathbf{r}}{\sqrt{(2\pi)^{3}}}4\pi\sum_{l^{\prime}m^{\prime}}i^{l^{\prime}}j_{l^{\prime}}(pr)Y_{l^{\prime}}^{m^{\prime}}(\hat{\mathbf{p}})\left(Y_{l^{\prime}}^{m^{\prime}}(\hat{\mathbf{r}})\right)^{*}\langle\mathbf{r}|\psi_{nlm}\rangle\\
 & = & \sqrt{\frac{2}{\pi}}i^{l}Y_{l}^{m}(\hat{\mathbf{p}})\int dr\left(r^{2}j_{l}(pr)R_{ln}(r)\right).
\end{eqnarray}
To find the radial Schr\"{o}dinger's equation for the Weyl coupling
in position space,
\begin{equation}
\left[-\frac{\nabla_{\mathbf{p}}^{2}}{2}+\frac{\mathbf{p}^{2}}{2}+v\mathbf{p}\cdot\boldsymbol{\sigma}\right]\psi_{njm}(\mathbf{p})=E_{nj}\psi_{njm}(\mathbf{p})
\end{equation}
 it is easiest to begin with the same equation in momentum space,
and Fourier transform the eigenstates $\psi_{njm}(\mathbf{p})=\int e^{i\mathbf{p}\cdot\mathbf{r}}\phi_{njm}(\mathbf{r)}d\mathbf{r}$.
But recall, we have showed that $j$ and $m$ are good quantum numbers,
and the eigenfunctions have the form $\psi_{njm}(\mathbf{p})=f_{n}^{+}(p)\chi_{jm}^{+}(\mathbf{\hat{p}})+f_{n}^{-}(p)\chi_{jm}^{-}(\mathbf{\hat{p}})$.
The spinors $\chi_{jm}^{\pm}$ contain spherical harmonics of order
$l=j\mp\frac{1}{2}$, so they are eigenfunctions of the Fourier transform.
The position space wavefunction thus has the form 
\begin{eqnarray}
\phi_{njm}(\mathbf{r}) & = & 4\pi\int\frac{d^{3}\mathbf{r}}{\sqrt{(2\pi)^{3}}}\left[\sum_{lm}(-i)^{l}j_{l}(pr)Y_{l}^{m}(\hat{\mathbf{p}})\left(Y_{l}^{m}(\hat{\mathbf{r}})\right)^{*}\right]\left(f_{n}^{+}(p)\chi_{jm}^{+}(\mathbf{\hat{p}})+f_{n}^{-}(p)\chi_{jm}^{-}(\mathbf{\hat{p}})\right)\\
 & = & \sqrt{\frac{2}{\pi}}(-i)^{j-\frac{1}{2}}\left(-ig_{n}^{+}(r)\chi_{jm}^{+}(\mathbf{\hat{r}})+g_{n}^{-}(r)\chi_{jm}^{-}(\mathbf{\hat{r}})\right),
\end{eqnarray}
where $g_{n}^{\pm}(r)=\int p^{2}j_{l}\left(pr\right)f_{n}^{\pm}(p)dp$,
and $l=j\mp\frac{1}{2}$. 

We are now ready to transform the momentum-space Schr\"{o}dinger
equation, we first multiply by $e^{i\mathbf{p}\cdot\mathbf{r}}$,
and then integrate over momentum to get 
\begin{eqnarray}
\int\frac{d^{3}\mathbf{p}}{\sqrt{(2\pi)^{3}}}e^{i\mathbf{p}\cdot\mathbf{r}}\left[-\frac{\nabla_{\mathbf{p}}^{2}}{2}+\frac{\mathbf{p}^{2}}{2}+v\mathbf{p}\cdot\boldsymbol{\sigma}\right]\psi_{njm}(\mathbf{p}) & = & E_{nj}\int\frac{d^{3}\mathbf{p}}{\sqrt{(2\pi)^{3}}}e^{i\mathbf{p}\cdot\mathbf{r}}\psi_{njm}(\mathbf{p}).
\end{eqnarray}
Transforming this term-by-term, the momentum space kinetic term becomes
a position space trapping term 
\begin{eqnarray}
\int\frac{d^{3}\mathbf{p}}{\sqrt{(2\pi)^{3}}}e^{i\mathbf{p}\cdot\mathbf{r}}\left(\frac{-\nabla_{\mathbf{p}}^{2}}{2}\right)\psi_{njm}(\mathbf{p}) & = & \frac{1}{2}\mathbf{r}^{2}\phi_{njm}(\mathbf{r}),
\end{eqnarray}
Similarly, the momentum space trap becomes a kinetic energy term in
position space 
\begin{eqnarray}
\int\frac{d^{3}\mathbf{p}}{\sqrt{(2\pi)^{3}}}\left(\frac{\mathbf{p}^{2}}{2}\right)e^{i\mathbf{p}\cdot\mathbf{r}}\psi_{njm}(\mathbf{p}) & = & -\frac{\nabla^{2}}{2}\int\frac{d^{3}\mathbf{p}}{\sqrt{(2\pi)^{3}}}\left(\psi_{njm}(\mathbf{p})e^{i\mathbf{p}\cdot\mathbf{r}}\right)\\
 & = & -\frac{\nabla^{2}}{2}\phi_{njm}(\mathbf{r}).
\end{eqnarray}
The spin-orbit term is more complicated, we first use the property
that $\mathbf{p}\cdot\boldsymbol{\sigma}\chi_{jm}^{\pm}(\hat{\mathbf{p}})=p\chi_{jm}^{\mp}(\hat{\mathbf{p}})$
\begin{eqnarray}
\int\frac{d^{3}\mathbf{p}}{\sqrt{(2\pi)^{3}}}\left[e^{i\mathbf{p}\cdot\mathbf{r}}v\left(\mathbf{p}\cdot\boldsymbol{\sigma}\right)\psi_{njm}(\mathbf{p})\right] & = & v\int\frac{d^{3}\mathbf{p}}{\sqrt{(2\pi)^{3}}}\left[e^{i\mathbf{p}\cdot\mathbf{r}}p\left(f_{n}^{+}(p)\chi_{jm}^{-}(\mathbf{\hat{p}})+f_{n}^{-}(p)\chi_{jm}^{+}(\mathbf{\hat{p}})\right)\right],
\end{eqnarray}
we then expand the exponential $e^{i\mathbf{p}\cdot\mathbf{r}}$ in
terms of spherical harmonics,
\begin{eqnarray}
\int\frac{d^{3}\mathbf{p}}{\sqrt{(2\pi)^{3}}}\left[e^{i\mathbf{p}\cdot\mathbf{r}}v\left(\mathbf{p}\cdot\boldsymbol{\sigma}\right)\psi_{njm}(\mathbf{p})\right] & = & v\frac{4\pi}{\sqrt{(2\pi)^{3}}}\sum_{lm}i^{l}\int d\hat{\mathbf{p}}\int dp\, p^{3}\left[j_{l}\left(pr\right)Y_{l}^{m}\left(\hat{\mathbf{r}}\right)\left(Y_{l}^{m}\left(\hat{\mathbf{p}}\right)\right)^{*}\left(f_{n}^{+}(p)\chi_{jm}^{-}(\mathbf{\hat{p}})+f_{n}^{-}(p)\chi_{jm}^{+}(\mathbf{\hat{p}})\right)\right]\\
 & = & v\frac{4\pi}{\sqrt{(2\pi)^{3}}}i^{j-\frac{1}{2}}\left[i\int dp\left(p^{3}j_{j+\frac{1}{2}}(pr)f_{nj}^{-}(p)\right)\chi_{jm}^{+}(\hat{\mathbf{r}})+\int dp\left(p^{3}j_{j-\frac{1}{2}}(pr)f_{nj}^{+}(p)\right)\chi_{jm}^{-}(\hat{\mathbf{r}})\right].
\end{eqnarray}
The spinors $\chi_{jm}^{\pm}$ have angular momentum components $l=j\mp\frac{1}{2}$.
These angular momentum variables do not match the angular momentum
corresponding to the spinors multiplying them. This, along with the
extra factor of $p$ in the integrand$\int dp\left(p^{3}j_{l}(pr)f(p)\right)=\int dp\, p^{2}\left(pj_{l}(pr)f(p)\right)$,
suggest it would be helpful to express the term $pj_{j_{\pm\frac{1}{2}}}(pr)$
in terms of spherical Bessel functions of the order $j\mp\frac{1}{2}$.
We use the identities 
\begin{eqnarray}
zj_{l}(z) & = & z\frac{d}{dz}j_{l+1}(z)+(l+2)j_{l+1}(z)\\
zj_{l}(z) & = & -z\frac{d}{dz}j_{l-1}(z)+(l-1)j_{l-1}(z)
\end{eqnarray}
to express 
\begin{eqnarray}
pj_{l}(pr) & = & \left(\frac{d}{dr}+\frac{l+2}{r}\right)j_{l+1}(pr)\\
 & = & \left(-\frac{d}{dr}+\frac{l-1}{r}\right)j_{l-1}(pr).
\end{eqnarray}
These relations convert the product $pj_{l}(pr)$ into a differential
operator in $r$ that can be removed from the integral. Together,
these give the expression for the Fourier transform of the three-dimensional
spin-orbit coupling term
\begin{eqnarray}
\int\frac{d^{3}\mathbf{p}}{\sqrt{(2\pi)^{3}}}\left[e^{i\mathbf{p}\cdot\mathbf{r}}v\left(\mathbf{p}\cdot\boldsymbol{\sigma}\right)\psi_{njm}(\mathbf{p})\right] & = & v\sqrt{\frac{2}{\pi}}i^{j-\frac{1}{2}}\left[i\left(-\frac{d}{dr}+\frac{j-\frac{1}{2}}{r}\right)\int dp\left(p^{2}j_{j-\frac{1}{2}}(pr)f_{nj}^{-}(p)\right)\chi_{jm}^{+}(\hat{\mathbf{r}})\right.\\
 & + & \left.\left(\frac{d}{dr}+\frac{j+\frac{1}{2}}{r}\right)\int dp\left(p^{3}j_{j-\frac{1}{2}}(pr)f_{nj}^{+}(p)\right)\chi_{jm}^{-}(\hat{\mathbf{r}})\right]\\
 & = & vi^{j-\frac{1}{2}}\left[i\left(-\frac{d}{dr}+\frac{j-\frac{1}{2}}{r}\right)g_{nj}^{-}(r)\chi_{jm}^{+}(\hat{\mathbf{r}})+\left(\frac{d}{dr}+\frac{j+\frac{1}{2}}{r}\right)g_{nj}^{+}(r)\chi_{jm}^{-}(\hat{\mathbf{r}})\right].
\end{eqnarray}
Finally, we combine these to get the expression for the radial Schr\"{o}dinger
equation for the three-dimensional spin-orbit coupling in position
space, 
\begin{eqnarray}
\frac{1}{2}\left[-\frac{1}{r}\left(\frac{d^{2}}{dr^{2}}r\right)+\frac{(j+\frac{1}{2})(j+\frac{3}{2})}{r^{2}}+r^{2}\right]g_{nj}^{+}(r)+v\left(-\frac{d}{dr}+\frac{j-\frac{1}{2}}{r}\right)g_{nj}^{-}(r) & = & E_{nj}g_{nj}^{+}(r)\\
\frac{1}{2}\left[-\frac{1}{r}\left(\frac{d^{2}}{dr^{2}}r\right)+\frac{(j+\frac{1}{2})(j-\frac{1}{2})}{r^{2}}+r^{2}\right]g_{nj}^{-}(r)+v\left(\frac{d}{dr}+\frac{j+\frac{1}{2}}{r}\right)g_{nj}^{+}(r) & = & E_{nj}g_{nj}^{-}(r).
\end{eqnarray}
\end{widetext}
\end{document}